\documentclass[journal=jacsat,manuscript=article]{achemso}

\usepackage{chemformula} 
\usepackage[T1]{fontenc} 
\setkeys{acs}{usetitle = true, maxauthors = 0}
\usepackage{color}
\usepackage{amsmath}
\usepackage{graphicx} 
\usepackage{bm}
\usepackage{cleveref}
\usepackage{longtable}
\usepackage{titlecaps}
\usepackage{listings}
\usepackage{multirow}
\usepackage{xr}
\usepackage{subfiles}
\usepackage{array}
\author{Marina Macchiagodena}
\affiliation{Dipartimento di Chimica ``Ugo Schiff'',
  Universit\`a degli Studi di Firenze, Via della Lastruccia 3, Sesto Fiorentino, I-50019 Italy} 
\author{Marco Pagliai}
\affiliation{Dipartimento di Chimica ``Ugo Schiff'',
  Universit\`a degli Studi di Firenze, Via della Lastruccia 3, Sesto Fiorentino, I-50019 Italy} 
\author{Piero Procacci}
\email{procacci@unifi.it} 
\affiliation{Dipartimento di Chimica ``Ugo Schiff'',
  Universit\`a degli Studi di Firenze, Via della Lastruccia 3, Sesto Fiorentino, I-50019 Italy} 

\title{Inhibition of the Main Protease 3CL$^{\rm Pro}$ of the Coronavirus
  Disease 19 via Structure-Based Ligand Design and Molecular
  Modeling.}

\abbreviations{COVID-19, 3CL$^{\rm Pro}$, Coronavirus inhibitors, Docking,
  ligand design, playmolecule, main protease, autodock}

\begin{document}
\begin{abstract}
We have applied a computational strategy, based on the synergy of
virtual screening, docking and molecular dynamics techniques, aimed at
identifying possible lead compounds for the non-covalent inhibition of
the main protease 3CL$^{\rm pro}$ of the SARS-Cov2 Coronavirus.  Based
on the recently resolved 6LU7 PDB structure, ligands were generated
using a multimodal structure-based design and then optimally docked to
the 6LU7 monomer. Docking calculations show that ligand-binding is
strikingly similar in SARS-CoV and SARS-CoV2 main proteases,
irrespectively of the protonation state of the catalytic CYS-HIS
dyad. The most potent docked ligands are found to share a common
binding pattern with aromatic moieties connected by rotatable bonds in
a pseudo-linear arrangement. Molecular dynamics calculations fully
confirm the stability in the 3CL$^{\rm pro}$ binding pocket of the most
potent binder identified by docking, namely a
chlorophenyl-pyridyl-carboxamide derivative.
\end{abstract}
At the beginning of this year, the world was dismayed by the outbreak
of a new severe viral acute respiratory syndrome (SARS), currently
known as COVID-19, that rapidly spreads from its origin in the Hubei
Chinese district to virtually whole China and, as of today, to more
than thirty nations in five continents.\cite{DONG2020} The new
coronavirus, named SARS-CoV2 and believed to have a zoonotic origin,
has infected thus far about 80000 people worldwide with nearly 10000
in critical conditions, causing the death of more than 3000
people. The SARS-CoV2's genome\cite{viralzone,ncbi} has a large
identity\cite{Shanker2020} with that of the SARS-CoV whose epidemic
started in early in 2003 and ended in the summer of the same year.

Most of the Coronaviridae genome encodes two large polyproteins, pp1a
and, through ribosomal frameshifting during
translation\cite{Thiel2003}, pp1ab. These polyproteins are cleaved and
transformed in mature non-structural proteins (NSPs) by the two
proteases 3CL$^{\rm pro}$ (3C-like protease) and PL$^{\rm pro}$
(Papain Like Protease) encoded by the open reading frame
1.\cite{Hilgenfeld2014} NSPs, in turn, play a fundamental role in the
transcription/replication during the infection.\cite{Thiel2003}
Targeting these proteases may hence constitute a valid approach for
antiviral drug design.  The catalytically active 3CL$^{\rm pro}$ is a
dimer. Cleavage by 3CL$^{\rm pro}$ occurs at the glutamine residue in
the P1 position of the substrate via the protease CYS-HIS dyad in
which the cysteine thiol functions as the nucleophile in the
proteolytic process.\cite{Anand2003} While dimerization is believed to
provide a substrate-binding cleft between the two
monomers,\cite{Chuck2010} in the dimer the solvent-exposed CYS-HYS
dyads are symmetrically located at the opposite edges the cleft,
probably acting independently.\cite{Shi4620} As no host-cell proteases
are currently known with this specificity, early drug discovery was
directed towards the so-called covalent Michael
inhibitors,\cite{Johansson2012} via electrophilic attack to the
cysteinate of the 3CL$^{\rm pro}$ dyad. On the other hand, the
consensus in drug discovery leads to excluding electrophiles from drug
candidates for reasons primarily relating to safety and adverse
effects such as allergies, tissue destruction, or
carcinogenesis.\cite{Vasudevan2019}

In spite of the initial effort in developing small-molecule compounds
(SMC) with anti-coronavirus activity immediately after the SARS
outbreak,\cite{Yang2003} no anti-viral drug was ever approved or even
reached the clinical stage due to a sharp decline in funding of
coronavirus research after 2005-2006, based on the erroneous
conviction by policy-makers and scientists that chance of a repetition
of a new zoonotic transmission was extremely unlikely. The most potent
non-covalent inhibitor for 3CL$^{\rm pro}$, ML188, was 
reported nearly ten years ago\cite{Jacobs2010} with moderate activity
in the low micromolar range.\cite{Jacobs2013}

\begin{figure}
    \centering
    \includegraphics[scale=0.40]{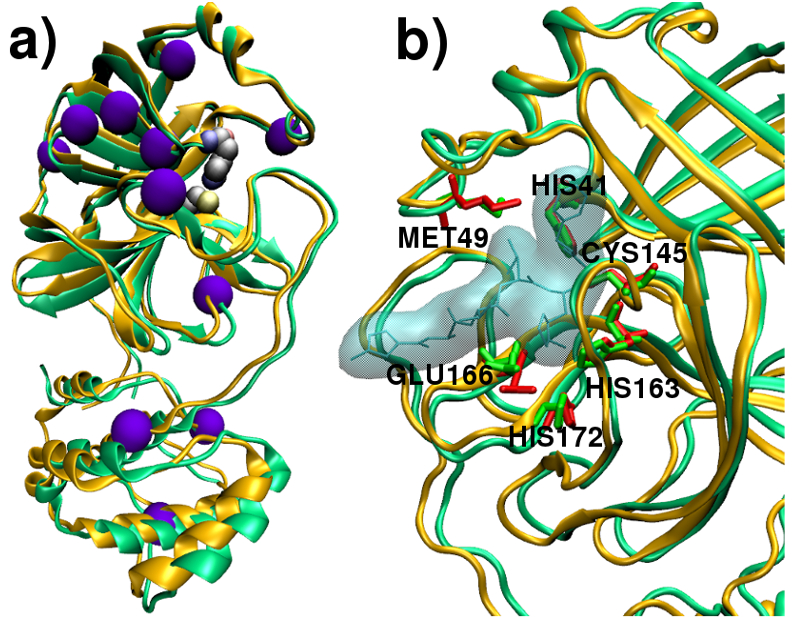}
    \caption{a):SARS-CoV2(orange, pdbcode 6LU7) and SARS-CoV (green,
      pdbcode 1UK4) main proteases. Violet spheres corresponds to the
      alpha carbons of the 12 differing residues in the two
      structures. Grey spheres indicate the CYS-HIS dyad b):
      view of the binding pocket with the main residues in
      bond representation (green and red for SARS-CoV2 and SARS-CoV,
      respectively). The shaded region mark the binding site for the
      substrate}
    \label{fig:res}
\end{figure}

According to the latest report of the structure of 3CL$^{\rm pro}$
from SARS-CoV2\cite{6lu7} (PDB code 6LU7) and the available structure
of 3CL$^{\rm pro}$ from SARS-CoV,\cite{Yang2003} (PDB code 1UK4), the
two main proteases differ by only 12 amino acids, with $\alpha$ carbon
atoms all lying at least 1 nm away from the 3CL$^{\rm pro}$ active
site (see Figure \ref{fig:res}a).
The substrate-binding pockets of two coronavirus main proteases are
compared in Figure \ref{fig:res}b, exhibiting a strikingly high level
of alignment of the key residues involved in substrate binding,
including the CYS145$\cdots$HIS41 dyad, and HIS163/HIS172/GLU166. The
latter residues are believed to provide the opening gate for the
substrate in the active state of the protomer.\cite{Yang2003}

Figure \ref{fig:res}(a,b) strongly suggest that effective non-covalent
inhibitors for SARS-CoV and SARS-CoV2 main proteases should share the
same structural and chemical features. In order to investigate this
matter, we have performed a molecular modeling study on both the 6LU7
and 1UK4 PDB structures. 6LU7 is the monomer of the main protease in
the active state with the N3 peptidomimetic inhibitor\cite{6lu7} while
1UK4 is the dimer with the protomer chain A in the active
state.\cite{Yang2003} The main protease monomer contains three
domains.  Domains I and II (residues 8-101 and residues 102-184) are
made of antiparallel $\beta$-barrel structures in a chymotrypsin-like
fold responsible for catalysis.\cite{Hu2009}

The 6LU7 structure was first fed to the {\em PlayMolecule} web
application\cite{play} using a novel virtual screening technique for
the multimodal structure-based ligand design\cite{skalic2019target},
called Ligand Generative Adversarial Network (LIGANN). Ligands in
LIGANN are generated so as to match the shape and chemical attributes
of the binding pocket and decoded into a sequence of SMILES enabling
directly the structure-based {\em{de novo}} drug design. SMILES codes
for ligands were obtained using the default LIGANN values for shapes
and channels with the cubic box center set at the midpoint vector
connecting the SH and NE atoms of the CYS-HIS dyad in the 6LU7 PDB
structure. The {\em PlayMolecule} interface delivered 93 optimally fit
non-congeneric compounds, spanning a significant portion of the
chemical space, whose SMILES and structures are reported in the
Supporting Information (SI).
\begin{figure}
    \centering
    \includegraphics[scale=0.7]{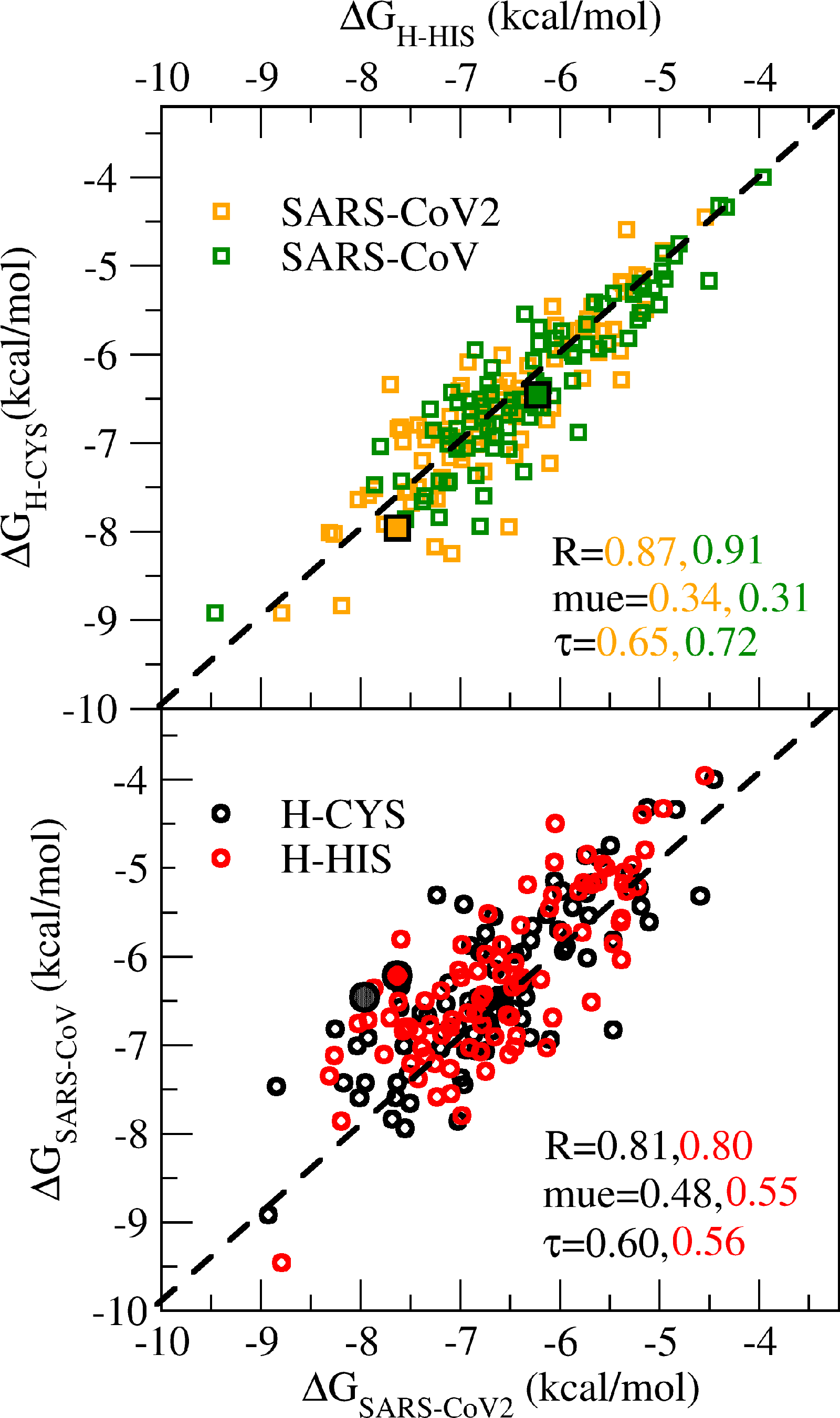}
    \caption{Correlation diagrams of autodock-computed binding free
      energies for 93 ligands of the SARS-CoV and
      SARS-CoV2 3CL$^{\rm pro}$ structures. $R,~ {\rm mue},~\tau$
      indicate the Pearson correlation coefficient, the mean unsigned
      error, and the Kendall rank coefficient, respectively. Upper
      panel: correlation diagram between ligand free energies obtained
      with the charged CYS$^{-1}$-HIS$^{+}$ (x-axis) and with neutral
      CYS-HIS (y-axis) dyad. Lower panel: correlation diagram between
      ligand free energies of SARS-CoV2 and SARS-CoV.}
    \label{fig:corr}
\end{figure}
 Each of these compounds was docked to the 6LU7 and to the 1UK4
 structures, using Autodock4\cite{Autodock} with full ligand
 flexibility. For both structures, the docking was repeated by setting
 the dyad with the residue in their neutral (CYS-HIS) and charged
 state (CYS$^-$/HIS$^+$).  Further details on Docking parameters are
 given in the SI.

Results for the binding free energies of the 93 LIGGAN-determined
3CL$^{\rm pro}$ ligands are reported in Figure \ref{fig:corr}. Binding
free energies are comprised in the range 4-9 kcal/mol and are found to
be strongly correlated for the two protonation states of the CYS-HIS
dyad. Correlation is still high when ligand binding free
energies for the main proteases are compared, confirming that good
binders for SARS-CoV are, in general, also good binders for SARS-CoV2
3CL$^{\rm pro}$.
\begin{figure}[H]
    \centering
    \includegraphics[scale=0.57]{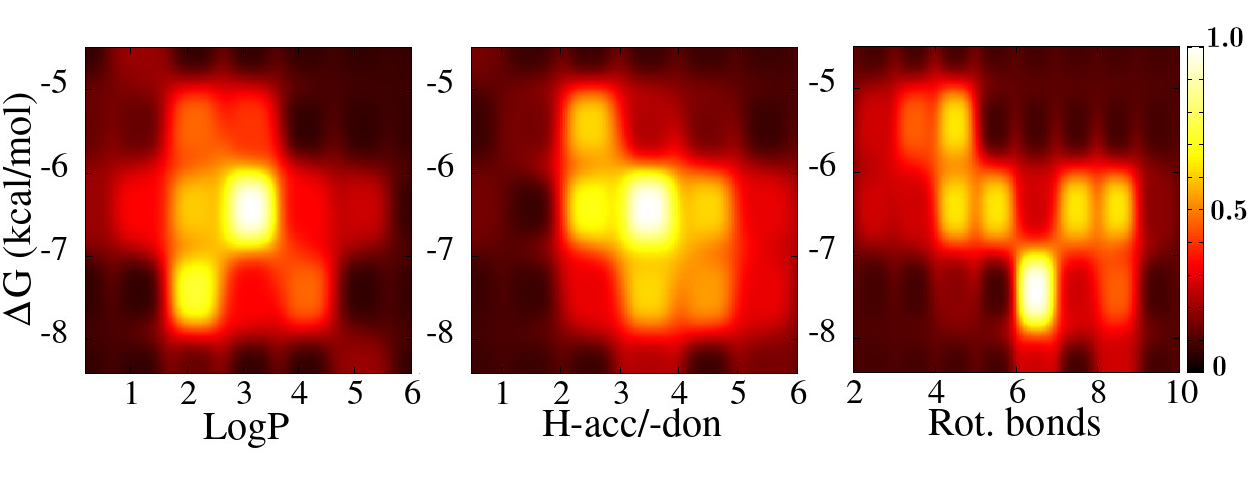}
    \caption{2D probability histograms $\Delta G$ with LogP (left),
      Hbond acceptors or donors (center) and rotatable bonds (right)
      for the LIGGAN-determined compounds of Table 1 of the Supporting
      Information. The common color-coded $z$-scale on the right
      corresponds to the 2D probability.}
    \label{fig:lpdg}
\end{figure}
For each of these compounds, using the knowledge-based XLOGP3
methodology\cite{Cheng2007}, we computed the octanol/water partition
coefficient (LogP) to assess the distribution in hydrophobic and
cytosolic environments. LogP values range from -0.5 to a
maximum of 5 with a number of rotatable bonds from 2 to a maximum of
12. Most of the LIGGAN compounds bear from 2 to 5 H-bond acceptor or
donors (see Table 1 of the SI). In Figure
\ref{fig:lpdg} we show the 2D probability distributions for $\Delta G$
correlated in turn to the LogP, number of H-bond donor/acceptors and
number of rotatable bonds.  We note, on the left and central panel,
sharp maxima for ${\rm LogP=3:4}$, $\Delta G=-7:-8$ and for
$\textit{\rm H-acc/don}=3$, $\Delta G=-6:-7$, respectively, suggestive
of a ligand-protein association driven mostly by hydrophobic
interactions. We must stress here that the computed $\Delta G$
pertains to the associations of the ligand with {\it one} protein
whatever the state of association of the protein. At free ligand
concentration equal to $K_d \equiv e^{-\Delta G/RT}$, i.e. when half
of the protein molecules are inhibited, the probability to have {\it
  both} monomers inhibited is equal to 1/4, whatever the dissociation
constant of the dimer,\cite{Graziano2006} hence the need for
identifying nanomolar or subnanomolar inhibitors of 3CL$^{\rm pro}$.

\begin{figure}
    \centering
    \includegraphics[scale=0.7]{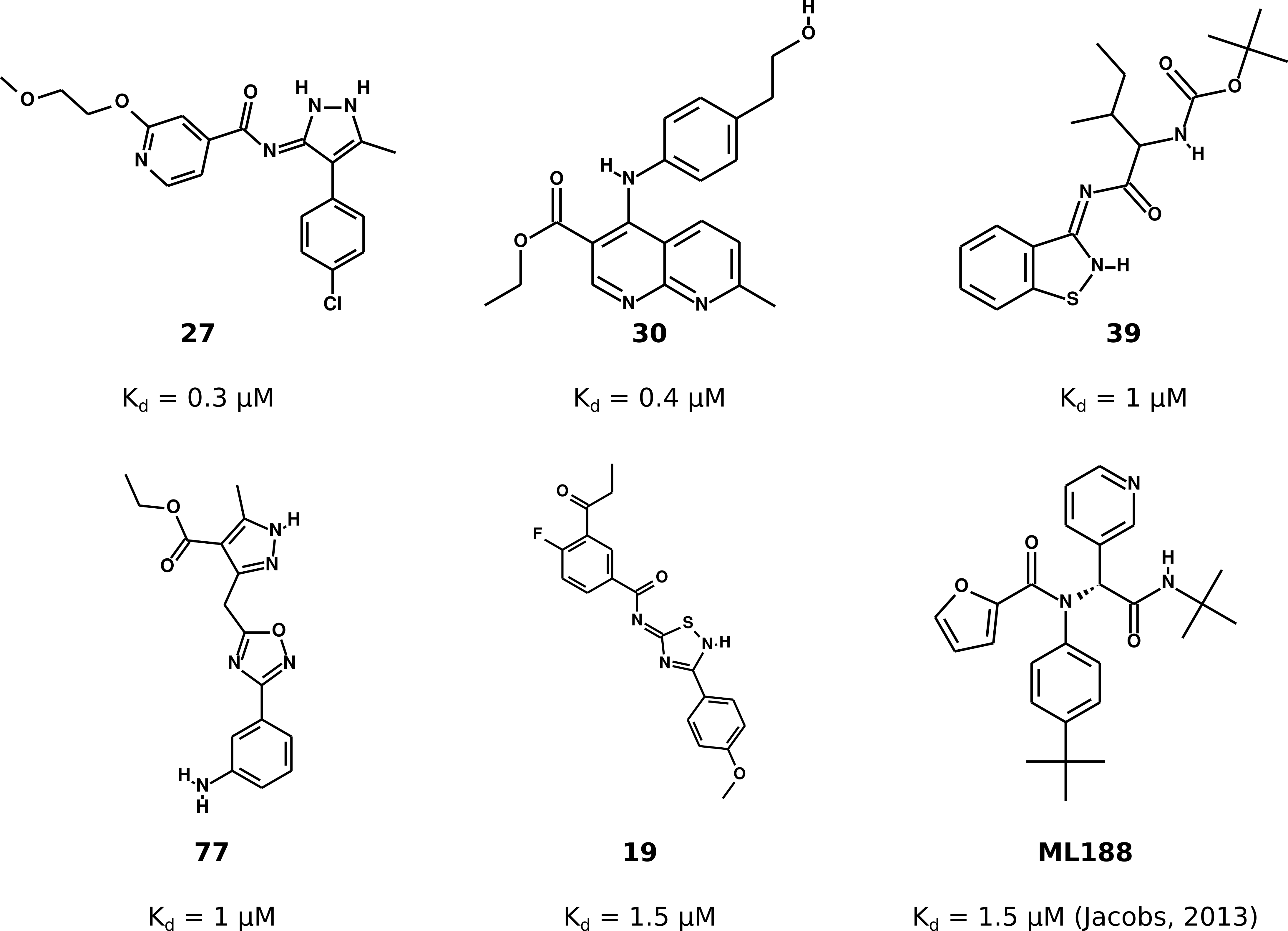}
    \caption{fig:Best binders for SARS-CoV2 main protease}
\label{fig:structures}
\end{figure}

Figure \ref{fig:structures} shows the chemical structures of the five
compounds exhibiting the highest binding affinity to the 6LU7 main
protease of SARS-CoV2 when the CYS-HIS dyad is in the neutral
state. None of these compounds is commercially available, although
some of them ({\bf 27}, {\bf 31}, {\bf 40}) show a high degree of
similarity with known structures according to the Tanimoto
metrics.\cite{pubchem} The LIGGAN-determined structures of Figure
\ref{fig:structures}, as well as many of those reported in Figures 1-5
of the SI, seem to share a common pattern with
aromatic moieties connected by rotatable bonds in a pseudo-linear
arrangement.  In Table \ref{tab:dg}, the binding free energy data of
these five best ligands are shown for both CoV proteases and both
protonation states of the catalytic dyad.
Inspection of Table \ref{tab:dg} confirms that SARS-CoV2 best binders
{\bf 27}, {\bf 29}, {\bf 39}, {\bf 77}, {\bf 19} are also good binders
for SARS-CoV 3CL$^{\rm pro}$. Remarkably, compound {\bf{27}} is
consistently the most potent ligand for the two proteases,
irrespective of the dyad protonation state. In the Table
\ref{tab:dg} we also report the Autodock4-computed binding free energy
for ML188. The Autodock4-predicted binding free energy for the
association of ML188-SARS-Cov protease is -6.2 and -6.5 kcal/mol for
the H-HIS and H-CYS tautomers, not too distant from the
experimentally determined value of -8 kcal/mol, hence lending support
for the LIGGAN-Autodock4 protocol used in identifying the lead
compounds of Table \ref{tab:dg}.
\begin{table}
  \begin{tabular} {l|rr|rr|r}
    \hline\hline
 & \multicolumn{2}{c|}{\bf CoV19}& \multicolumn{2}{c|}{\bf SARS} &   \\ \hline
        {Comp.} & H-CYS & H-HIS & H-CYS & H-HIS & LogP\\ \hline
        27 & -8.92 & -8.79   &  -8.92  &  -9.46  &    4.90\\
30 & -8.84 & -8.19   &  -7.47  &  -7.86  &    3.74\\
39 & -8.25 & -7.08   &  -6.82  &  -6.72  &    6.06\\
77 & -8.17 & -7.25   &  -7.43  &  -7.21  &    2.03\\
19 & -8.03 & -8.26   &  -7.01  &  -7.12  &    5.58\\
ML188 & -7.96 & 7.63 & 6.46 & 6.22$^a$  & 4.97 \\
    \hline\hline
  \end{tabular}
\caption{Computed binding free energies (kcal/mol), $\Delta G$, of the
  best five binders (see the full list in the Supporting information)
  for SARS-CoV2 3CL$^{\rm pro}$. $\Delta G$ values are reported for
  the two protonation state of the dyad and for SARS-Cov and SARS-Cov2
  main protease.}
 {$^a$Experimental value for ML188 is\cite{Jacobs2013}
   $\Delta G=-7.98$ kcal/mol.}
  \label{tab:dg}
\end{table}

In order to assess the stability of the 3CL$^{\rm pro}$-{\bf 27}
association, we have performed extensive molecular dynamics
simulations\cite{gromacs,gromacs1} of the bound state with explicit
solvent. The overall structural information was obtained by combining
data from three independent simulations (for a total of about 120 ns),
all started from the best docking pose of {\bf 27} on the 6LU7
monomeric structure. Further methodological
aspects\cite{macchiagodena2019} are provided in the Supporting
Information.
\begin{figure}
    \centering
    \includegraphics[scale=0.7]{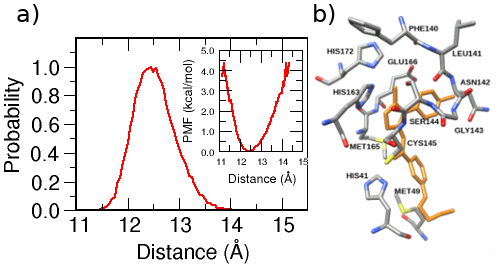}
    \caption{a) Probability distribution of the distance between the
      centers of mass of compound {\bf 27} and domain I+II of
      3CL$^{\rm PRO}$ as obtained from triplicates MD simulations (120
      ns in total) (in the inset the corresponding PMF is shown). b)
      Binding pocket of 6LU7 with ligand {\bf 27}. The time record of the
      minimal distances between ligand and the depicted nearby
      residues are reported in Figures 6, 7 of the Supporting
      Information.}
    \label{fig:pmfpocket}
\end{figure}
In Figure \ref{fig:pmfpocket}, we show the probability distribution,
$P(R)$, of the distance $R$ between the center of mass (CoM) of the
ligand and that of the domains I+II. The distribution has nearly a
Gaussian shape with a half-width of about 1 \AA, exhibiting only a
minor positive skewness and defining a tight binding site volume
$V_{\rm site}$ of few \AA$^{3}$~at most.\cite{statbind} The
MD-determined $P(R)$ shows that the ligand never leaves the binding
pocket at any stage during the whole simulation.  In the inset of
Figure \ref{fig:pmfpocket}a, we show the potential of mean force (PMF)
along the ligand-protein CoM distance $R$, computed as
$v(R)=-RT\log(P(R)/\max[P(r)])$. As $1/K_d$ = $ \int_{V_{\rm site}}
e^{-\beta v(R)} d{\bf R}$,\cite{Gilson1997}, the steepness of the
curve is suggestive of a profound minimum and hence of a large
association constant, confirming the indication obtained from the
Docking calculations.  Figure \ref{fig:pmfpocket}b shows polar and
hydrophobic residues found in at least 90\% of the simulation time
within 4.5 and 5.5 \AA~ distance, respectively, from any atom of the
ligand {\bf 27}. All essential residues for binding are included, with
the important addition of Met165, Phe140 and Leu141 hydrophobic
residues that consistently linger near the pyrazolic or the
chlorinated phenyl rings of {\bf 27}, in agreement with the
hydrophobic character of the interaction.

Figures \ref{fig:lpdg} and \ref{fig:pmfpocket} shows possible avenues
for improvement. For example, forcing the L-shaped binding structure
(see Figure \ref{fig:pmfpocket}b) in bulk also, possibly by
redesigning the rotatable connectors in the ligand, may reduce the
penalty due conformational entropy loss upon binding,\cite{statbind}
hence boosting the ligand affinity for 3CL$^{\rm pro}$. Building upon
this knowledge, we hence plan to optimize the lead using MD
simulations coupled to efficient relative binding free energy
calculation via free energy perturbation on congeneric variants
\cite{Shirts2013}, eventually providing {\em in silico} determined
anti-viral compounds to be synthesized an experimentally tested {\em
  in vitro} and {\em in vivo}.

While the infection rate for COVID 19 in China is currently declining
for days, new shocking outbreaks are developing in the South Corea,
the Middle East and Europe, with high risk for a pandemic. The
scientific community is hence called to an extraordinary and
collaborative effort for a rapid delivering of an effective anti-COVID
19 drug, hoping that our contribution can be of help in such a
worldwide endeavor.

\providecommand{\latin}[1]{#1}
\makeatletter
\providecommand{\doi}
  {\begingroup\let\do\@makeother\dospecials
  \catcode`\{=1 \catcode`\}=2 \doi@aux}
\providecommand{\doi@aux}[1]{\endgroup\texttt{#1}}
\makeatother
\providecommand*\mcitethebibliography{\thebibliography}
\csname @ifundefined\endcsname{endmcitethebibliography}
  {\let\endmcitethebibliography\endthebibliography}{}


\clearpage

\title{\bf Supporting Information for ``Inhibition of the Main Protease 3CL$^{\rm pro}$ of the Coronavirus
  Disease 19 via Structure-Based Ligand Design and Molecular
  Modeling''}

\vspace{2cm}
\setcounter{figure}{0}
\setcounter{table}{0}

\tiny{
\begin{longtable}{l|c|c|c|c|c|c|c|c|c}
\caption{SMILES code and labeling used in the paper for ligands obtained using LIGGAN.}\\
\hline
 & && & &&\multicolumn{4}{c}{\bf{$\Delta$G (kcal/mol)}} \\\cline{7-10}
\bf{SMILES} & \bf{Comp.}&\bf{logP}&H$_{\rm acc}$ & H$_{\rm don}$&$n_{\rm rot}$ &\multicolumn{2}{c|}{\bf{SARS-CoV-2}}&\multicolumn{2}{c}{\bf{SARS-CoV}} \\\cline{7-10}
&&&&&&H-CYS&H-HID&H-CYS&H-HID\\
\hline\hline
\endfirsthead
 & && & &&\multicolumn{4}{c}{\bf{$\Delta$G (kcal/mol}} \\\cline{7-10}
\bf{SMILES} & \bf{Comp.}&\bf{logP}&H$_{\rm acc}$ & H$_{\rm don}$&$n_{\rm rot}$ &\multicolumn{2}{c|}{\bf{SARS-CoV-2}}&\multicolumn{2}{c}{\bf{SARS-CoV}} \\\cline{7-10}
&&&&&&H-CYS&H-HID&H-CYS&H-HID\\
\hline\hline
\endhead
NC(=O)c1csc(CNCC(N)C(F)(F)F)c1  &    \bf{1} &             0.120  &  3 &   1 &   5  &            -5.46  &        -5.68  &        -6.83  &        -6.52  \\
CCC(C(=O)NC(CC(C)C)C(=O)O)(C(=O)c1ccccc1)C(C)C  &    \bf{2} &             4.450  &  2 &   4 &  11  &    -6.64  &        -6.49  &        -6.15  &        -6.68  \\
COCCc1ccc(Oc2ccc(Nc3nc(C)cc(C)n3)cc2)cc1        &    \bf{3} &             4.520  &  1 &   2 &   7  &    -7.56  &        -7.55  &        -7.01  &        -6.80  \\
CCc1cc(=O)oc2ccc(-c3cccc4c3CCOC4)cc12   &    \bf{4} &             2.870  &  0 &   1 &   2  &    -6.34  &        -7.70  &        -6.46  &        -6.69  \\
CCOC(=O)C(NC(=O)Cc1c[nH]c2cccc(C)c12)C(C)(C)C   &    \bf{5} &             3.670  &  1 &   2 &   8  &    -6.35  &        -6.99  &        -6.30  &        -6.24  \\
COCCn1ccc2c(C(=O)Nc3ccc4c(c3)nnn4C)cccc21       &    \bf{6} &             1.810  &  1 &   3 &   6  &    -7.32  &        -6.76  &        -6.67  &        -6.43  \\
COc1cccc(-c2cc(O)cc(F)n2)c1     &    \bf{7} &             2.560  &  1 &   2 &   3  &    -5.10  &        -5.22  &        -5.61  &        -5.21  \\
CCN(CCC1CC1)C(=O)C1CCCN(CCNC(C)=O)C1    &    \bf{8} &             1.370  &  1 &   2 &  10  &    -6.46  &        -6.19  &        -6.07  &        -6.26  \\
CCOC(=O)c1c(NCc2c[nH]nc2C)sc(CC)c1-c1ccc(F)cc1  &    \bf{9} &             5.440  &  1 &   2 &   8  &    -7.95  &        -6.51  &        -7.43  &        -7.11  \\
CC(C)N1CC(c2ncc(-c3cc(Br)c[nH]c3=O)[nH]2)CC1=O  &    \bf{10} &            2.150  &  0 &   3 &   3  &    -6.74  &        -6.74  &        -5.74  &        -5.98  \\
CCC(NC(=O)Cc1c(C)[nH]c2ccccc12)C(CC)S(=O)(=O)C1CC1      &    \bf{11} &            3.510  &  1 &   3 &   9  &    -7.20  &        -7.38  &        -7.03  &        -7.03  \\
CC(C)(C)C(C)(C)CCCNC(=S)NCC(N)=O        &    \bf{12} &            2.540  &  3 &   1 &   9  &    -6.27  &        -5.77  &        -5.66  &        -5.73  \\
Cc1cc(F)cc2[nH]c(C(=O)Nc3cncc(C(=O)OC(C)(C)C)c3)cc12    &    \bf{13} &            3.450  &  1 &   3 &   6  &    -6.84  &        -7.62  &        -7.07  &        -6.51  \\
O=C(Nc1ccc2ccoc2c1)c1sc2cc(C(F)(F)F)ccc2c1F     &    \bf{14} &            5.500  &  1 &   1 &   3  &    -6.13  &        -6.32  &        -5.53  &        -5.19  \\
CC(=O)C(NC(=O)c1cnc2c(c1)c(=O)[nH]c(=O)n2C1CC1)C(C)C    &    \bf{15} &            3.020  &  1 &   5 &   6  &    -6.37  &        -6.61  &        -5.95  &        -6.04  \\
COCC(C)CC(=O)Nc1ncc(Cc2cc(F)cc(F)c2)s1  &    \bf{16} &            3.200  &  1 &   2 &   8  &    -6.74  &        -6.13  &        -7.07  &        -7.03  \\
COCCn1cc(C(=O)N=c2[nH][nH]c(C)c2-c2ccc(F)cc2)cn1        &    \bf{17} &            3.380  &  0 &   2 &   6  &    -6.68  &        -6.90  &        -6.84  &        -7.03  \\
COc1ccc(-c2cccc3c2COC3=O)cc1    &    \bf{18} &            2.940  &  0 &   1 &   2  &    -5.73  &        -5.68  &        -5.20  &        -5.19  \\
CCC(=O)c1cc(C(=O)N=c2nc(-c3ccc(OC)cc3)[nH]s2)ccc1F      &    \bf{19} &            5.580  &  0 &   3 &   6  &    -8.03  &        -8.26  &        -7.01  &        -7.12  \\
CCc1ccc(NC(=O)c2ccc3ccccc3c2)cc1F       &    \bf{20} &            4.900  &  1 &   1 &   4  &    -7.23  &        -6.10  &        -5.31  &        -5.46  \\
COc1ccc(Oc2ncccn2)cc1OC &    \bf{21} &            2.010  &  0 &   2 &   4  &    -5.74  &        -5.56  &        -4.86  &        -4.96  \\
COCCOc1ccc(NC(=O)c2sc(-c3cnn(C)c3)nc2C)cc1      &    \bf{22} &            2.320  &  1 &   3 &   8  &    -6.91  &        -7.19  &        -6.51  &        -6.39  \\
CCC(c1ccc(F)cc1)c1nc(-c2ccc(CNC)cc2)no1 &    \bf{23} &            3.910  &  1 &   2 &   6  &    -7.02  &        -7.09  &        -7.86  &        -7.55  \\
CCOC(=O)COc1ccc(NC(=O)c2c[nH]cc2C)cc1Cl &    \bf{24} &            2.800  &  1 &   2 &   8  &    -7.63  &        -7.23  &        -7.43  &        -7.59  \\
COc1cc2c(cc1F)oc1ccc(OCC(F)F)cc12       &    \bf{25} &            4.560  &  0 &   0 &   4  &    -5.98  &        -5.61  &        -5.26  &        -5.16  \\
COCC(=O)Nc1cccc(-c2cnn(Cc3nc(C)cs3)c2)c1F       &    \bf{26} &            2.080  &  1 &   3 &   7  &    -6.57  &        -6.45  &        -6.47  &        -6.07  \\
COCCOc1cc(C(=O)N=c2[nH][nH]c(C)c2-c2ccc(Cl)cc2)ccn1     &    \bf{27} &            4.900  &  0 &   2 &   7  &    -8.92  &        -8.79  &        -8.92  &        -9.46  \\
CC(=O)NCCN(CC1CCCCC1)CC1CCCOC1  &    \bf{28} &            2.550  &  1 &   1 &   8  &    -6.89  &        -6.72  &        -5.88  &        -5.52  \\
CC(C(=O)N1CC2CCC1C2C(=O)O)=C1C(=O)c2ccccc2SC1=S &    \bf{29} &            2.730  &  1 &   4 &   4  &    -6.61  &        -6.07  &        -6.44  &        -6.69  \\
CCOC(=O)c1cnc2nc(C)ccc2c1Nc1ccc(CCO)cc1 &    \bf{30} &            3.740  &  2 &   4 &   8  &    -8.84  &        -8.19  &        -7.47  &        -7.86  \\
COCC(=O)Nc1ccc(-c2noc(-c3cncc(Cl)c3)n2)cc1      &    \bf{31} &            2.080  &  1 &   4 &   6  &    -7.55  &        -7.51  &        -7.94  &        -6.80  \\
COCc1occc1-c1nc2cc(F)ccc2o1     &    \bf{32} &            2.280  &  0 &   1 &   3  &    -5.29  &        -5.27  &        -5.06  &        -4.97  \\
COCCc1ccc2cc(Nc3nnc(-c4cnn(C)c4)n3C)ccc2c1      &    \bf{33} &            2.640  &  1 &   3 &   6  &    -6.94  &        -7.08  &        -7.06  &        -6.93  \\
COC(=O)CC(NC(=O)c1c[nH]cc1-c1cnn(C)c1)c1ccncc1  &    \bf{34} &            0.210  &  1 &   4 &   8  &    -6.80  &        -7.42  &        -5.95  &        -6.85  \\
COc1cccc(-c2nc3cccnc3s2)c1F     &    \bf{35} &            3.380  &  0 &   2 &   2  &    -5.60  &        -5.72  &        -4.89  &        -4.85  \\
CCOC(=O)C(CC)(CC)CNC(=O)C(c1ccccc1)C1CC1        &    \bf{36} &            4.120  &  1 &   2 &  11  &    -7.14  &        -6.45  &        -6.54  &        -7.02  \\
COc1cc(-c2cccc3nsnc23)sn1       &    \bf{37} &            2.820  &  0 &   3 &   2  &    -5.93  &        -5.97  &        -5.89  &        -5.73  \\
Cc1[nH]c2ccccc2c(=O)c1CC(=O)N(CC(=O)OC(C)(C)C)C(C)C     &    \bf{38} &            4.360  &  0 &   3 &   8  &    -7.64  &        -8.02  &        -7.60  &        -6.76  \\
CCC(C)C(NC(=O)OC(C)(C)C)C(=O)N=c1[nH]sc2ccccc12 &    \bf{39} &            6.060  &  1 &   2 &   8  &    -8.25  &        -7.08  &        -6.82  &        -6.72  \\
COCCn1c(C)c(C)c2cc(NC(=O)c3c[nH]cc3-c3ccc(C)o3)ccc21    &    \bf{40} &            3.300  &  1 &   1 &   7  &    -7.19  &        -6.98  &        -7.04  &        -7.80  \\
CNCc1cccc(NC2CCCCNC2)c1 &    \bf{41} &            1.760  &  3 &   0 &   4  &    -6.09  &        -6.92  &        -6.94  &        -6.64  \\
COc1ccsc1-c1ccc(C(=O)O)c(F)c1   &    \bf{42} &            2.860  &  1 &   2 &   4  &    -5.12  &        -5.17  &        -4.32  &        -4.40  \\
CCc1cccc(-n2c(=O)[nH]c3ccc(OC(C)=O)cc3c2=O)c1   &    \bf{43} &            5.120  &  0 &   3 &   4  &    -6.57  &        -6.79  &        -6.52  &        -6.48  \\
COc1ccc2c(c1)OCCO2      &    \bf{44} &            1.610  &  0 &   0 &   1  &    -4.45  &        -4.54  &        -4.00  &        -3.96  \\
COCCOc1ncccc1C(=O)N=c1cc(-c2cccc(F)c2F)[nH][nH]1        &    \bf{45} &            4.100  &  0 &   2 &   7  &    -6.70  &        -7.10  &        -6.51  &        -6.80  \\
COC(=O)C(C)c1cccc(NC(=O)c2cnn(-c3ccncc3)c2)n1   &    \bf{46} &            1.320  &  1 &   5 &   7  &    -6.70  &        -6.43  &        -6.76  &        -6.90  \\
COc1ccc(-c2cscc2C(=O)O)cc1      &    \bf{47} &            2.730  &  1 &   2 &   4  &    -5.29  &        -5.35  &        -5.25  &        -5.05  \\
CCCC(C(=O)O)N(Cc1ccc(-n2cncn2)c(F)c1)c1nn[nH]n1 &    \bf{48} &            2.140  &  1 &   7 &   9  &    -5.96  &        -5.39  &        -5.94  &        -5.61  \\
CCOC(=O)c1nc(Cn2ccc(-c3ccc4ncccc4c3)c2N)no1     &    \bf{49} &            2.610  &  1 &   4 &   6  &    -7.68  &        -7.49  &        -7.84  &        -7.21  \\
COc1cccc(-c2nc(O)cc(O)n2)c1     &    \bf{50} &            1.790  &  2 &   4 &   4  &    -5.72  &        -5.46  &        -6.02  &        -5.86  \\
CCOC(=O)CCN(C(=O)c1[nH]c2ccccc2c1Br)C(C)(C)C    &    \bf{51} &            3.710  &  0 &   2 &   8  &    -7.02  &        -6.84  &        -6.79  &        -6.65  \\
CN(C)Cc1cccc(CN(C)C)c1  &    \bf{52} &            1.590  &  0 &   0 &   4  &    -4.59  &        -5.33  &        -5.32  &        -5.26  \\
CCOC(=O)c1c[nH]c2c(NCc3ccc(-n4ccnc4)nc3)ccnc12  &    \bf{53} &            1.780  &  1 &   4 &   7  &    -6.41  &        -7.01  &        -6.35  &        -6.16  \\
CCc1nc2ccc(C(=O)Nc3ccc(C)cc3C)cc2s1     &    \bf{54} &            4.780  &  1 &   2 &   4  &    -6.46  &        -6.38  &        -6.60  &        -6.24  \\
Cc1noc2nc(-c3ccc(F)cc3)cc(C(=O)NCCCCO)c12       &    \bf{55} &            2.440  &  2 &   4 &   8  &    -6.99  &        -7.57  &        -7.37  &        -6.84  \\
CCC1(C(=O)N=c2nc[nH]c3ccccc23)CCCN1C(=O)OC(C)(C)C       &    \bf{56} &            4.460  &  0 &   3 &   6  &    -6.96  &        -7.34  &        -7.45  &        -7.13  \\
COc1cccc(-c2cnc3ccc(F)cc3n2)c1  &    \bf{57} &            2.940  &  0 &   2 &   2  &    -6.05  &        -6.05  &        -5.15  &        -4.94  \\
COc1cc(C(F)(F)F)ccc1-c1cn2cccnc2n1      &    \bf{58} &            3.630  &  0 &   2 &   2  &    -5.49  &        -5.14  &        -4.75  &        -4.80  \\
COCCN(CC1CCC1)CC1CCN(C2CC2)CC1  &    \bf{59} &            2.740  &  0 &   0 &   8  &    -6.66  &        -6.47  &        -5.55  &        -6.35  \\
CC(C)CC(=O)Nc1c[nH]nc1-c1cc(-c2ccc(F)cc2)no1    &    \bf{60} &            2.560  &  1 &   3 &   6  &    -7.59  &        -7.92  &        -6.34  &        -6.72  \\
COc1ccc(-c2ccccc2OC)cc1 &    \bf{61} &            3.490  &  0 &   0 &   3  &    -5.20  &        -5.36  &        -5.23  &        -5.18  \\
N=C(N)c1cccc(NC2CCCC2)c1        &    \bf{62} &            2.140  &  2 &   0 &   3  &    -5.71  &        -5.76  &        -5.54  &        -5.17  \\
CCOC(=O)C(C)N1CC2CCCC(NCC(C)C)C2C1      &    \bf{63} &            3.050  &  1 &   1 &   7  &    -6.86  &        -7.35  &        -6.60  &        -6.50  \\
CC(C)(C)OC(=O)Nc1cc(-c2nc(-c3c[nH]c(=O)[nH]3)co2)co1    &    \bf{64} &            2.510  &  1 &   3 &   6  &    -7.17  &        -7.10  &        -6.86  &        -7.27  \\
COc1cc(-c2cccc3c2OCC3=O)ccc1O   &    \bf{65} &            2.720  &  1 &   2 &   3  &    -6.01  &        -6.58  &        -5.70  &        -6.21  \\
NCc1cccc(NC2CCCC(O)C2)c1        &    \bf{66} &            1.410  &  3 &   1 &   4  &    -6.54  &        -6.74  &        -6.62  &        -7.30  \\
CN1CCNC(c2cccc(CN)c2)C1=O       &    \bf{67} &           -0.350  &  2 &   1 &   2  &    -6.93  &        -6.82  &        -7.06  &        -6.66  \\
CC(C(=O)Nc1ccc(C(C)(C)C)cc1)c1ccccc1F   &    \bf{68} &            5.040  &  1 &   1 &   5  &    -6.68  &        -6.44  &        -5.88  &        -6.21  \\
CCc1ccccc1NC(=O)c1ccc(-c2nnc(C)o2)cc1   &    \bf{69} &            3.440  &  1 &   3 &   5  &    -6.77  &        -6.65  &        -6.53  &        -6.91  \\
CC(C)(C)N1CCCC1C1CCN(C(=O)C2CC(N)C2)CC1 &    \bf{70} &            1.630  &  1 &   1 &   4  &    -6.30  &        -6.52  &        -6.92  &        -6.66  \\
COCCn1c(-c2cccc(-n3cccn3)c2)nc2c(=O)[nH]c(=O)n(C)c2c1=O &    \bf{71} &            3.360  &  0 &   5 &   5  &    -6.76  &        -6.78  &        -6.72  &        -6.77  \\
COCCNc1nc(-c2nc(-c3ccc4ncccc4c3)no2)cs1 &    \bf{72} &            2.970  &  1 &   4 &   6  &    -7.60  &        -7.30  &        -6.58  &        -6.77  \\
NC(=O)c1csc(CN2CCC(F)(F)CC2)c1  &    \bf{73} &            1.510  &  1 &   1 &   3  &    -5.67  &        -6.04  &        -5.17  &        -4.50  \\
COC(=O)C(C)(C)N(C)CCC1CCN(C(C)C)CC1     &    \bf{74} &            2.710  &  0 &   1 &   7  &    -7.11  &        -6.98  &        -6.30  &        -5.87  \\
C\#Cc1cccc(N(C)C(=O)c2ccc(-n3ccnc3)cc2)c1        &    \bf{75} &            2.940  &  0 &   2 &   5  &    -6.45  &        -6.58  &        -5.98  &        -5.87  \\
COc1ccccc1-c1ccc(CO)c(C=O)c1    &    \bf{76} &            2.100  &  1 &   2 &   5  &    -6.38  &        -6.41  &        -6.71  &        -6.30  \\
CCOC(=O)c1c(Cc2nc(-c3cccc(N)c3)no2)n[nH]c1C     &    \bf{77} &            2.030  &  1 &   4 &   6  &    -8.17  &        -7.25  &        -7.43  &        -7.21  \\
NC(=O)COc1ccc(-c2nc(-c3cccn4ccnc34)no2)cc1      &    \bf{78} &            2.220  &  1 &   4 &   5  &    -7.52  &        -7.86  &        -7.33  &        -6.36  \\
NCc1cccc(NC2(C(F)F)CC2)c1       &    \bf{79} &            1.990  &  2 &   0 &   4  &    -5.18  &        -5.38  &        -5.43  &        -5.57  \\
COCCc1nc2cc(NC(=O)c3cc4[nH]nnc4cc3C)ccc2o1      &    \bf{80} &            2.290  &  1 &   4 &   6  &    -8.01  &        -8.31  &        -7.60  &        -7.35  \\
NCCNc1cccc(CN2CCC2=O)c1 &    \bf{81} &           -0.030  &  2 &   1 &   5  &    -6.29  &        -5.38  &        -5.82  &        -6.04  \\
COCCn1ccc2cc(NC(=O)c3cnc4onc(C)c4c3)ccc21       &    \bf{82} &            2.070  &  1 &   3 &   6  &    -7.92  &        -7.76  &        -6.92  &        -7.11  \\
CCC(=O)c1cc(C(=O)Nc2ccc(-c3cnc[nH]3)cc2)ccc1OC  &    \bf{83} &            2.680  &  1 &   3 &   7  &    -6.82  &        -7.60  &        -6.46  &        -6.19  \\
COC(=O)C1=CCC(NC2CC(N)C23CCCCC3)C1      &    \bf{84} &            2.140  &  2 &   1 &   4  &    -7.39  &        -7.18  &        -6.64  &        -6.89  \\
CCC(=O)c1cc(C(=O)c2cc(C)n(-c3cccc(O)c3)c2C)ccc1F        &    \bf{85} &            4.360  &  1 &   3 &   6  &    -7.50  &        -7.42  &        -7.66  &        -7.38  \\
CC(=O)CC(C)(C)CC(=O)NC(c1c[nH]c2ccccc12)C(C)(C)C        &    \bf{86} &            3.700  &  1 &   2 &   8  &    -6.55  &        -6.79  &        -6.43  &        -7.08  \\
COc1ccc(-c2cnccn2)cc1OC &    \bf{87} &            1.390  &  0 &   2 &   3  &    -4.83  &        -4.96  &        -4.34  &        -4.33  \\
COCc1nc(-c2cccc(OC)c2)no1       &    \bf{88} &            1.460  &  0 &   2 &   4  &    -5.73  &        -5.81  &        -5.29  &        -5.26  \\
COCCC1(C(=O)NCC2CCN(CC(F)(F)F)CC2)CCC1  &    \bf{89} &            2.470  &  1 &   1 &   8  &    -6.96  &        -6.39  &        -5.41  &        -5.65  \\
COc1ccsc1-c1ccc2c(c1)CC(CO)O2   &    \bf{90} &            2.620  &  1 &   1 &   4  &    -5.87  &        -5.53  &        -5.44  &        -5.00  \\
CN(C)Cc1cccc(N=C(N)NCCO)c1      &    \bf{91} &           -0.040  &  3 &   1 &   7  &    -5.46  &        -6.07  &        -5.82  &        -5.31  \\
COCCOc1cc(Nc2ccc3c(C)nn(C)c3c2)ccc1O    &    \bf{92} &            2.980  &  2 &   2 &   7  &    -6.64  &        -6.82  &        -6.60  &        -6.17  \\
CS(=O)(=O)CC1CCCN(CC2CCCC2CN)C1 &    \bf{93} &            0.990  &  1 &   2 &   5  &    -6.86  &        -7.59  &        -6.88  &        -5.81  \\
\hline\hline
\end{longtable}
}
{\normalsize
\begin{figure}
    \centering
    \includegraphics[scale=0.8]{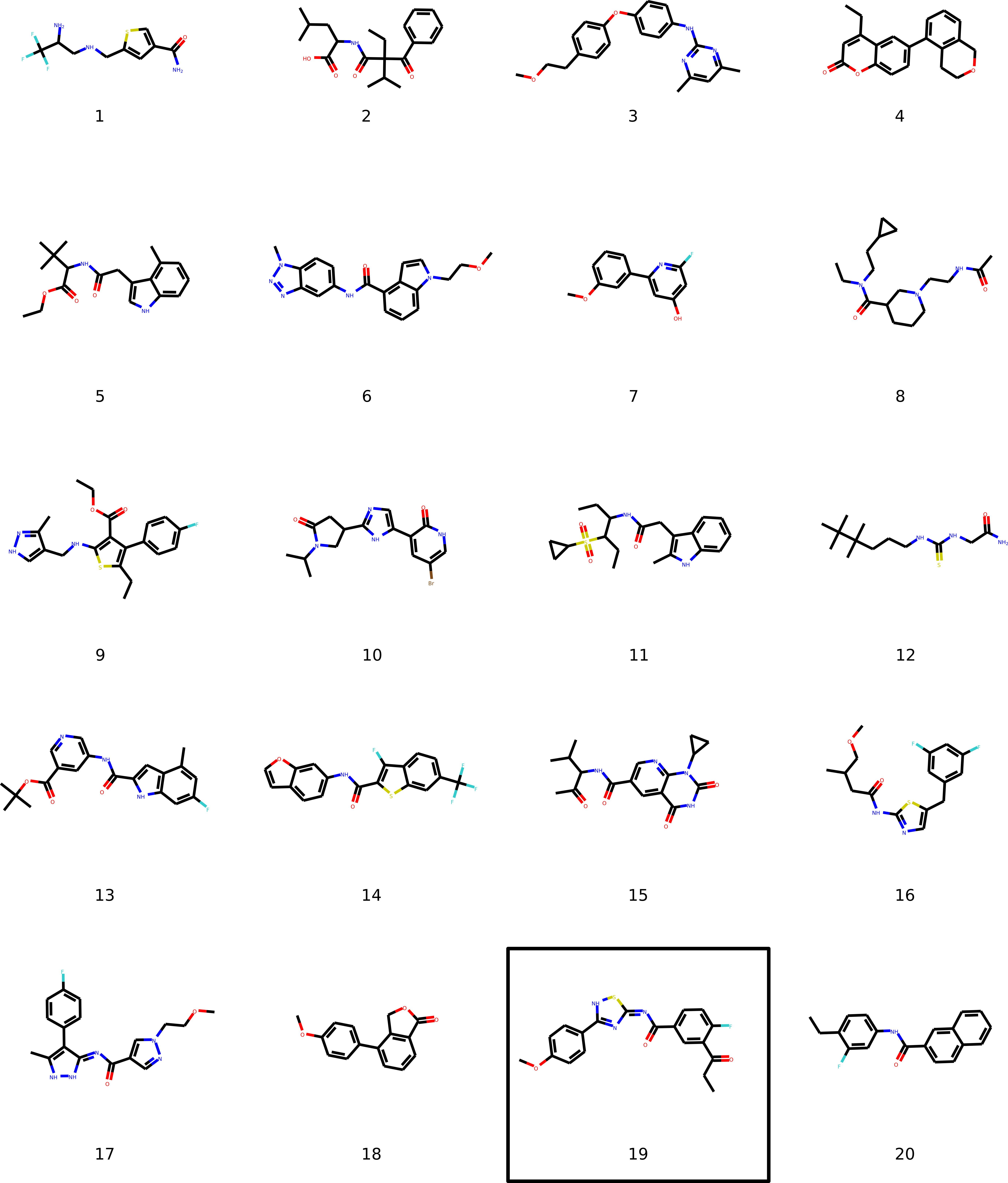}
    \caption{2D-structures for compounds 1 to 20. In box one of the best five binders.}
    \label{fig:120}
\end{figure}

\begin{figure}
    \centering
    \includegraphics[scale=0.8]{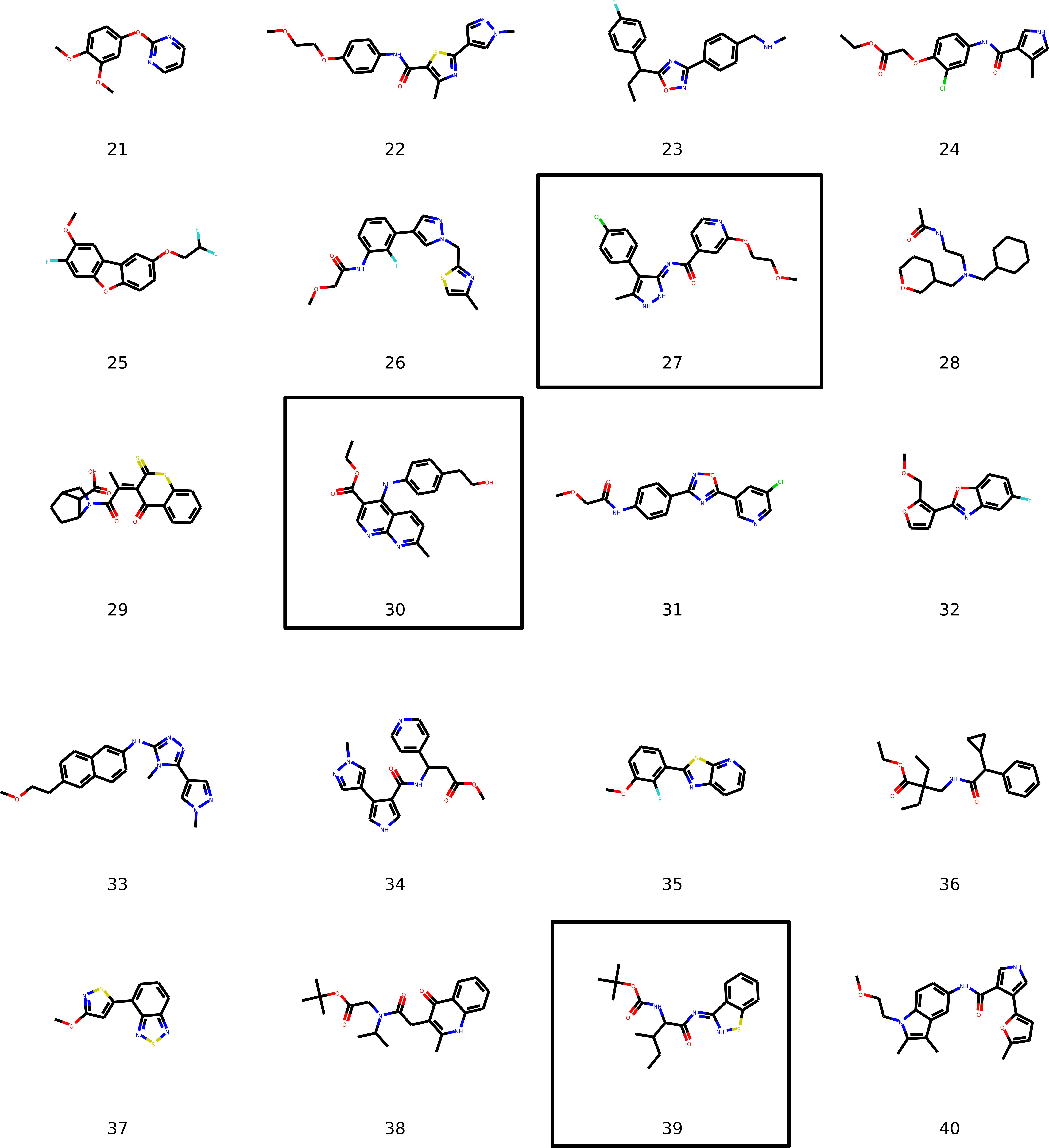}
    \caption{2D-structures for compounds 21 to 40. In box three of the best five binders.}
    \label{fig:2140}
\end{figure}

\begin{figure}
    \centering
    \includegraphics[scale=0.8]{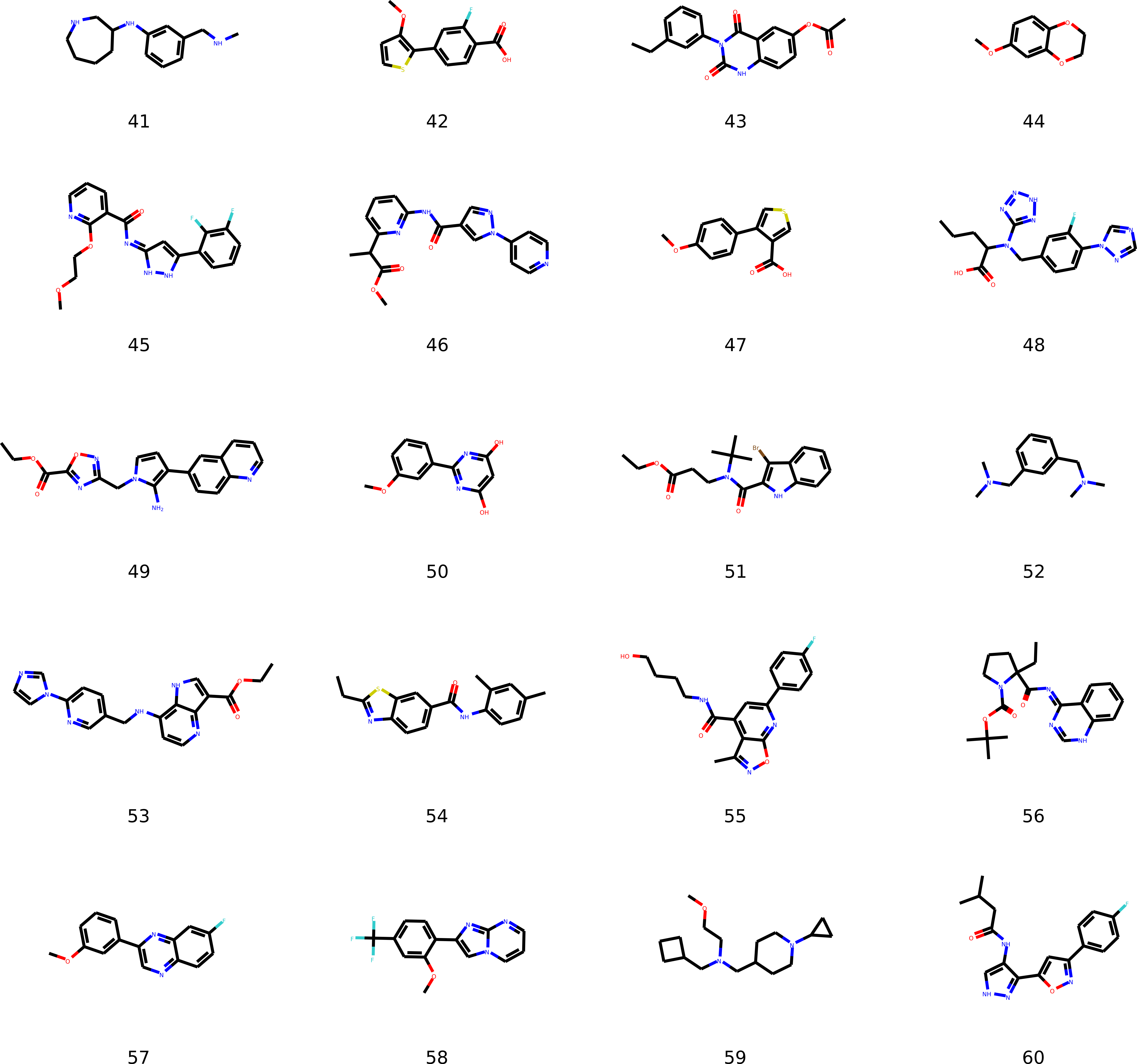}
    \caption{2D-structures for compounds 41 to 60.}
    \label{fig:4260}
\end{figure}

\begin{figure}
    \centering
    \includegraphics[scale=0.8]{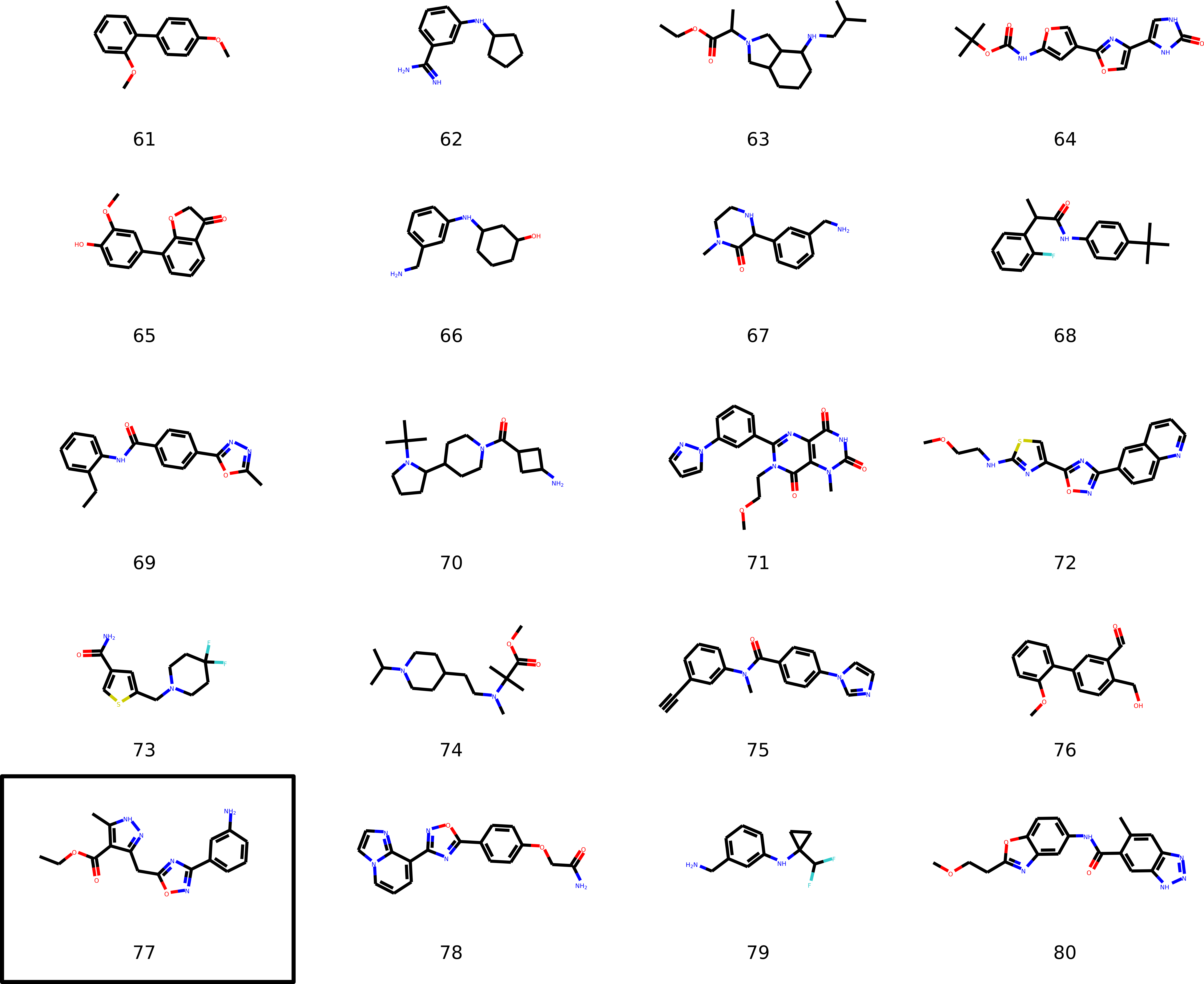}
    \caption{2D-structures for compounds 61 to 80. In box one of the best five binders.}
    \label{fig:6180}
\end{figure}

\begin{figure}
    \centering
    \includegraphics[scale=0.8]{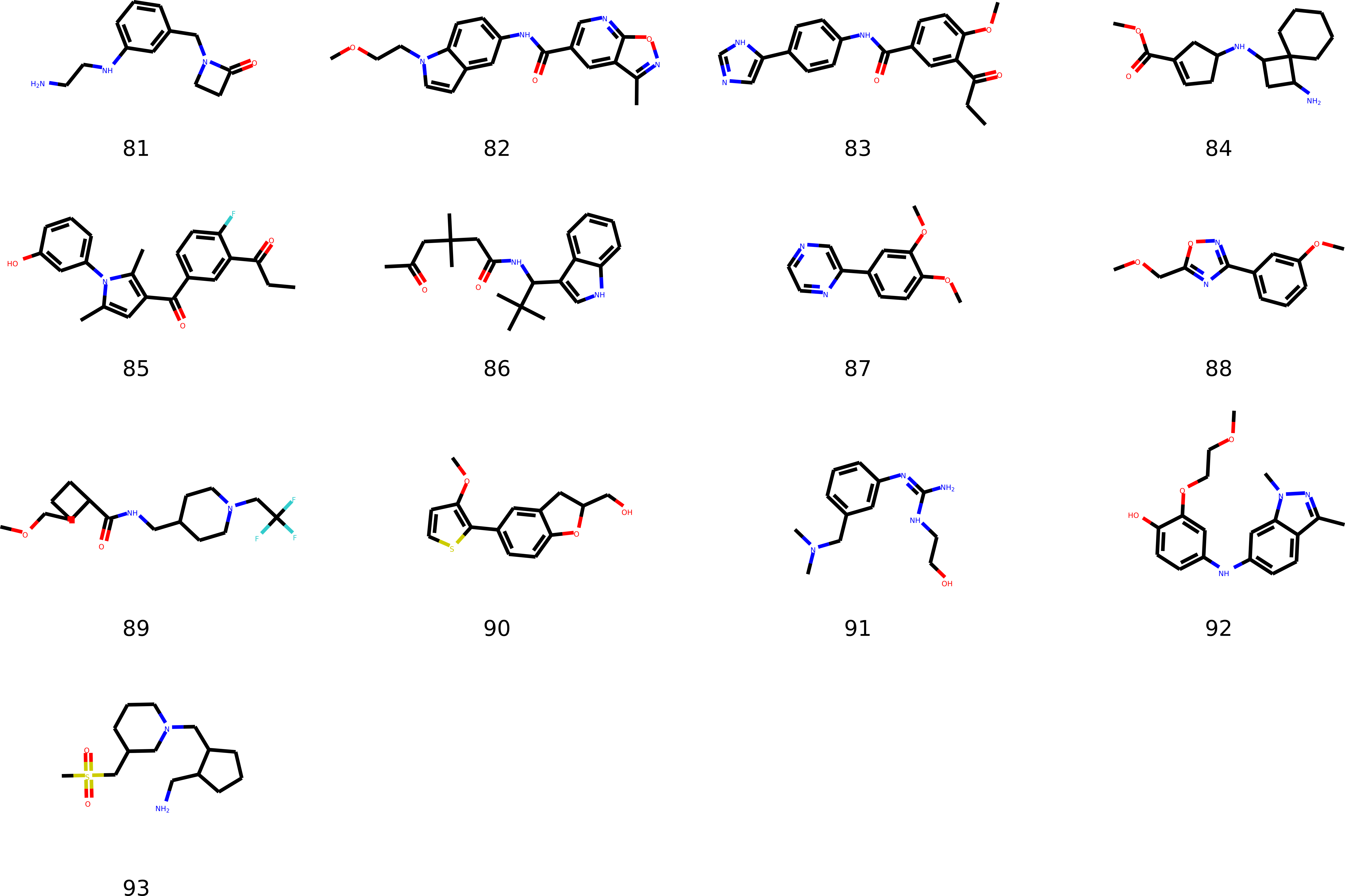}
    \caption{2D-structures for compounds 81 to 93.}
    \label{fig:8193}
\end{figure}

\clearpage
\section{Computational Details}
\subsection{Molecular Docking}
Following the two input file used for molecular docking.
\begin{itemize}
    \item file.gpf
    \begin{verbatim}
npts 60 60 60                        
gridfld 6lu7.maps.fld              
spacing 0.375                        
receptor_types A C HD N OA SA        
ligand_types ...              
receptor 6lu7.pdbqt                
gridcenter -10.18 20.65 66.75        
smooth 0.5                           
map 6lu7.A.map
map ...
...
elecmap 6lu7.e.map                
dsolvmap 6lu7.d.map              
dielectric -0.1465                  
\end{verbatim}
\item file.dpf
\begin{verbatim}
autodock_parameter_version 4.2       
outlev 1                            
intelec                              
seed pid time                       
ligand_types   ....          
fld 6lu7.maps.fld                  
map 6lu7.A.map
map ...
...
elecmap 6lu7.e.map              
desolvmap 6lu7.d.map           
move ligand.pdbqt                   
about 6.66645 -3.33598 -3.92332                  
tran0 random                         
quaternion0 random                   
dihe0 random                         
torsdof 4                            
rmstol 2.0                          
extnrg 1000.0                        
e0max 0.0 10000                     
ga_pop_size 150                     
ga_num_evals 2500000                 
ga_num_generations 27000             
ga_elitism 1                         
ga_mutation_rate 0.02               
ga_crossover_rate 0.8                
ga_window_size 10                   
ga_cauchy_alpha 0.0                  
ga_cauchy_beta 1.0                   
set_ga                              
unbound_model bound                  
do_global_only 50                   
analysis                            
\end{verbatim}

\end{itemize}

\subsection{Molecular Dynamics Simulations}
Molecular dynamics (MD) simulations were carried out in a cubic box with periodic boundary
conditions, whose side-length was chosen so that the minimum distance
between protein atoms belonging to neighboring replicas was larger
than 14 {\AA} in any direction. The system (protein+compound) was explicitly solvated with the SPC/E
\cite{spce} water model at the standard density. The starting configuration was generated using
GROMACS\cite{gromacs,gromacs1} and PrimadORAC.\cite{primadorac} 
The system was initially minimized at 0 K with a steepest
descent procedure and subsequently heated to 298.15 K in an NPT
ensemble (P=1 atm) using Berendsen barostat\cite{berendsen} and
velocity rescaling algorithm\cite{bussi} with an integration time step
of 0.1 fs and a coupling constant of 0.1 ps for 250 ps.

Production run in the NPT ensemble were carried out starting three independent simulations with different initial velocities randomization. Each MD run has been performed for 40 ns (for a total of 120 ns) imposing rigid
constraints \tcb{only on the X-H bonds (with X being any heavy atom)} by means
of the LINCS algorithm ($\delta$t=2.0 fs).\cite{lincs} Electrostatic
interactions were treated by using particle-mesh Ewald (PME)\cite{pme}
method with a grid spacing of 1.2 {\AA} and a spline interpolation of
order 4. The cross interactions for Lennard-Jones terms were calculated using the
Lorentz-Berthelot\cite{lorentz,berthelot} mixing rules and we 
excluded intramolecular non-bonded interactions between atom pairs separated \tcb{up
  to two bonds. The non-bonded interactions between 1-4 atoms involved
in a proper torsion were scaled by the standard AMBER fudge factors
(0.8333 and 0.5 for the Coulomb and Lennard-Jones, respectively).}

The simulations and the trajectories analysis were carried out using
the GROMACS 2018.3 program.\cite{gromacs,gromacs1}

\begin{figure}
    \centering
    \includegraphics[scale=1]{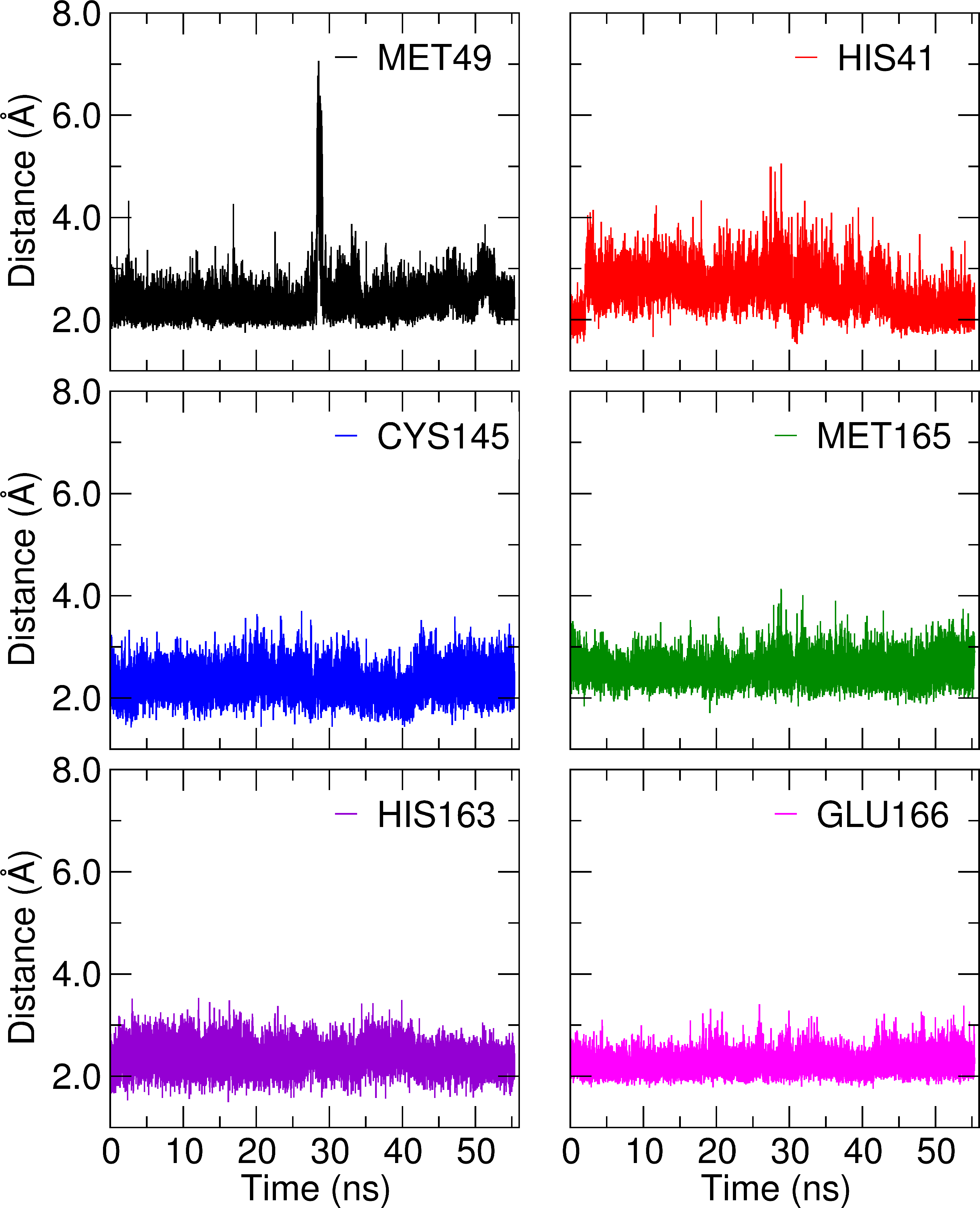}
    \caption{Pair distribution function between compound {\bf{27}} and pocket residues (MET49, HIS41, CYS145, MET165, HIS163, GLU166), calculated during 55 ns of MD simulation.}
    \label{fig:dist1}
\end{figure}

\begin{figure}
    \centering
    \includegraphics[scale=1]{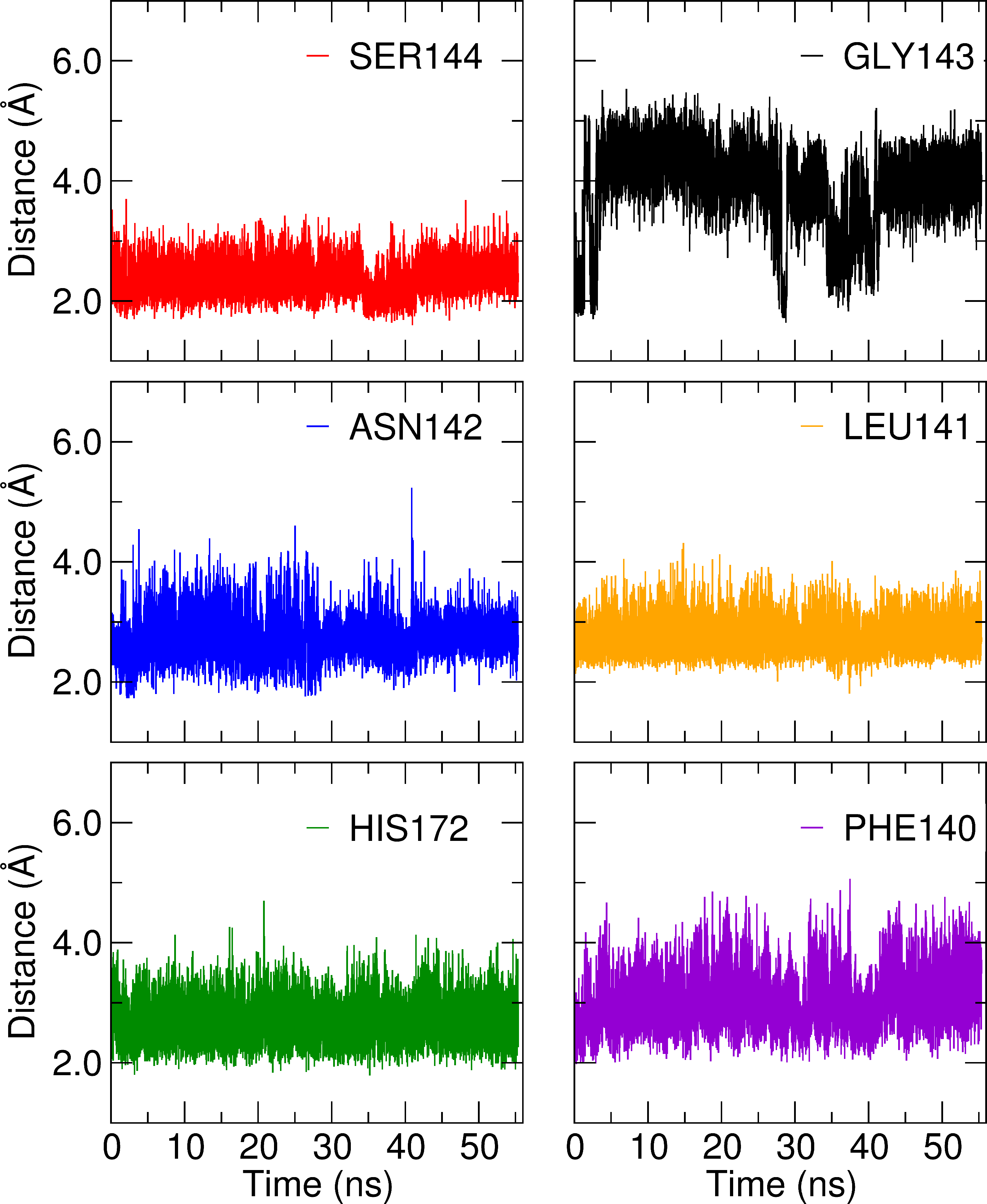}
    \caption{Pair distribution function between compound {\bf{27}} and pocket residues (SER144, GLY143, ASN142, LEU141, HIS172, PHE140), calculated during 55 ns of MD simulation.}
    \label{fig:dist2}
\end{figure}

}

\clearpage
\newpage
{\normalsize
\providecommand{\latin}[1]{#1}
\makeatletter
\providecommand{\doi}
  {\begingroup\let\do\@makeother\dospecials
  \catcode`\{=1 \catcode`\}=2 \doi@aux}
\providecommand{\doi@aux}[1]{\endgroup\texttt{#1}}
\makeatother
\providecommand*\mcitethebibliography{\thebibliography}
\csname @ifundefined\endcsname{endmcitethebibliography}
  {\let\endmcitethebibliography\endthebibliography}{}

}


\providecommand{\latin}[1]{#1}
\makeatletter
\providecommand{\doi}
  {\begingroup\let\do\@makeother\dospecials
  \catcode`\{=1 \catcode`\}=2 \doi@aux}
\providecommand{\doi@aux}[1]{\endgroup\texttt{#1}}
\makeatother
\providecommand*\mcitethebibliography{\thebibliography}
\csname @ifundefined\endcsname{endmcitethebibliography}
  {\let\endmcitethebibliography\endthebibliography}{}
\begin{mcitethebibliography}{11}
\providecommand*\natexlab[1]{#1}
\providecommand*\mciteSetBstSublistMode[1]{}
\providecommand*\mciteSetBstMaxWidthForm[2]{}
\providecommand*\mciteBstWouldAddEndPuncttrue
  {\def\EndOfBibitem{\unskip.}}
\providecommand*\mciteBstWouldAddEndPunctfalse
  {\let\EndOfBibitem\relax}
\providecommand*\mciteSetBstMidEndSepPunct[3]{}
\providecommand*\mciteSetBstSublistLabelBeginEnd[3]{}
\providecommand*\EndOfBibitem{}
\mciteSetBstSublistMode{f}
\mciteSetBstMaxWidthForm{subitem}{(\alph{mcitesubitemcount})}
\mciteSetBstSublistLabelBeginEnd
  {\mcitemaxwidthsubitemform\space}
  {\relax}
  {\relax}

\bibitem[Berendsen \latin{et~al.}(1987)Berendsen, Grigera, and Straatsma]{spce}
Berendsen,~H. J.~C.; Grigera,~J.~R.; Straatsma,~T.~P. The Missing Term in
  Effective Pair Potentials. \emph{J. Phys. Chem.} \textbf{1987}, \emph{91},
  6269--6271\relax
\mciteBstWouldAddEndPuncttrue
\mciteSetBstMidEndSepPunct{\mcitedefaultmidpunct}
{\mcitedefaultendpunct}{\mcitedefaultseppunct}\relax
\EndOfBibitem
\bibitem[Pronk \latin{et~al.}(2013)Pronk, Páll, Schulz, Larsson, Bjelkmar,
  Apostolov, Shirts, Smith, Kasson, van~der Spoel, Hess, and Lindahl]{gromacs}
Pronk,~S.; Páll,~S.; Schulz,~R.; Larsson,~P.; Bjelkmar,~P.; Apostolov,~R.;
  Shirts,~M.~R.; Smith,~J.~C.; Kasson,~P.~M.; van~der Spoel,~D.; Hess,~B.;
  Lindahl,~E. GROMACS 4.5: a High-Throughput and Highly Parallel Open Source
  Molecular Simulation Toolkit. \emph{Bioinformatics} \textbf{2013}, \emph{29},
  845\relax
\mciteBstWouldAddEndPuncttrue
\mciteSetBstMidEndSepPunct{\mcitedefaultmidpunct}
{\mcitedefaultendpunct}{\mcitedefaultseppunct}\relax
\EndOfBibitem
\bibitem[Van Der~Spoel \latin{et~al.}(2005)Van Der~Spoel, Lindahl, Hess,
  Groenhof, Mark, and Berendsen]{gromacs1}
Van Der~Spoel,~D.; Lindahl,~E.; Hess,~B.; Groenhof,~G.; Mark,~A.~E.;
  Berendsen,~H. J.~C. GROMACS: Fast, Flexible, and Free. \emph{J. Comput.
  Chem.} \textbf{2005}, \emph{26}, 1701--1718\relax
\mciteBstWouldAddEndPuncttrue
\mciteSetBstMidEndSepPunct{\mcitedefaultmidpunct}
{\mcitedefaultendpunct}{\mcitedefaultseppunct}\relax
\EndOfBibitem
\bibitem[Procacci(2017)]{primadorac}
Procacci,~P. PrimaDORAC: A Free Web Interface for the Assignment of Partial
  Charges, Chemical Topology, and Bonded Parameters in Organic or Drug
  Molecules. \emph{J. Chem. Inf. Model.} \textbf{2017}, \emph{57},
  1240--1245\relax
\mciteBstWouldAddEndPuncttrue
\mciteSetBstMidEndSepPunct{\mcitedefaultmidpunct}
{\mcitedefaultendpunct}{\mcitedefaultseppunct}\relax
\EndOfBibitem
\bibitem[Berendsen \latin{et~al.}(1984)Berendsen, Postma, van Gunsteren,
  Di~Nola, and Haak]{berendsen}
Berendsen,~H. J.~C.; Postma,~J. P.~M.; van Gunsteren,~W.~F.; Di~Nola,~A.;
  Haak,~J.~R. Molecular Dynamics with Coupling to an External Bath. \emph{J.
  Chem. Phys.} \textbf{1984}, \emph{81}, 3684--3690\relax
\mciteBstWouldAddEndPuncttrue
\mciteSetBstMidEndSepPunct{\mcitedefaultmidpunct}
{\mcitedefaultendpunct}{\mcitedefaultseppunct}\relax
\EndOfBibitem
\bibitem[Bussi \latin{et~al.}(2007)Bussi, Donadio, and Parrinello]{bussi}
Bussi,~G.; Donadio,~D.; Parrinello,~M. Canonical Sampling Through Velocity
  Rescaling. \emph{J. Chem. Phys.} \textbf{2007}, \emph{126}, 014101\relax
\mciteBstWouldAddEndPuncttrue
\mciteSetBstMidEndSepPunct{\mcitedefaultmidpunct}
{\mcitedefaultendpunct}{\mcitedefaultseppunct}\relax
\EndOfBibitem
\bibitem[Hess \latin{et~al.}(1997)Hess, Bekker, Berendsen, and Fraaije]{lincs}
Hess,~B.; Bekker,~H.; Berendsen,~H.; Fraaije,~J. LINCS: A Linear Constraint
  Solver for Molecular Simulations. \emph{J. Comput. Chem.} \textbf{1997},
  \emph{18}, 1463--1472\relax
\mciteBstWouldAddEndPuncttrue
\mciteSetBstMidEndSepPunct{\mcitedefaultmidpunct}
{\mcitedefaultendpunct}{\mcitedefaultseppunct}\relax
\EndOfBibitem
\bibitem[Darden \latin{et~al.}(1993)Darden, York, and Pedersen]{pme}
Darden,~T.; York,~D.; Pedersen,~L. Particle Mesh Ewald: An N log(N) Method for
  Ewald Sums in Large Systems. \emph{J. Chem. Phys.} \textbf{1993}, \emph{98},
  10089--10092\relax
\mciteBstWouldAddEndPuncttrue
\mciteSetBstMidEndSepPunct{\mcitedefaultmidpunct}
{\mcitedefaultendpunct}{\mcitedefaultseppunct}\relax
\EndOfBibitem
\bibitem[Antoon(1881)]{lorentz}
Antoon,~L.~H. Ueber die Anwendung des Satzes vom Virial in der Kinetischen
  Theorie der Gase. \emph{Ann. Phys} \textbf{1881}, \emph{248}, 127--136\relax
\mciteBstWouldAddEndPuncttrue
\mciteSetBstMidEndSepPunct{\mcitedefaultmidpunct}
{\mcitedefaultendpunct}{\mcitedefaultseppunct}\relax
\EndOfBibitem
\bibitem[Marcellin(1898)]{berthelot}
Marcellin,~B. Sur Le Mélange Des Gaz. \emph{Comptes Rendus Acad. Sci.}
  \textbf{1898}, \emph{126}, 1703--1855\relax
\mciteBstWouldAddEndPuncttrue
\mciteSetBstMidEndSepPunct{\mcitedefaultmidpunct}
{\mcitedefaultendpunct}{\mcitedefaultseppunct}\relax
\EndOfBibitem
\end{mcitethebibliography}


\providecommand{\latin}[1]{#1}
\makeatletter
\providecommand{\doi}
  {\begingroup\let\do\@makeother\dospecials
  \catcode`\{=1 \catcode`\}=2 \doi@aux}
\providecommand{\doi@aux}[1]{\endgroup\texttt{#1}}
\makeatother
\providecommand*\mcitethebibliography{\thebibliography}
\csname @ifundefined\endcsname{endmcitethebibliography}
  {\let\endmcitethebibliography\endthebibliography}{}
\begin{mcitethebibliography}{32}
\providecommand*\natexlab[1]{#1}
\providecommand*\mciteSetBstSublistMode[1]{}
\providecommand*\mciteSetBstMaxWidthForm[2]{}
\providecommand*\mciteBstWouldAddEndPuncttrue
  {\def\EndOfBibitem{\unskip.}}
\providecommand*\mciteBstWouldAddEndPunctfalse
  {\let\EndOfBibitem\relax}
\providecommand*\mciteSetBstMidEndSepPunct[3]{}
\providecommand*\mciteSetBstSublistLabelBeginEnd[3]{}
\providecommand*\EndOfBibitem{}
\mciteSetBstSublistMode{f}
\mciteSetBstMaxWidthForm{subitem}{(\alph{mcitesubitemcount})}
\mciteSetBstSublistLabelBeginEnd
  {\mcitemaxwidthsubitemform\space}
  {\relax}
  {\relax}

\bibitem[Dong \latin{et~al.}(2020)Dong, Du, and Gardner]{DONG2020}
Dong,~E.; Du,~H.; Gardner,~L. An Interactive Web-Based Dashboard to Track
  COVID-19 in Real Time. \emph{Lancet Infect. Dis.} \textbf{2020}, \relax
\mciteBstWouldAddEndPunctfalse
\mciteSetBstMidEndSepPunct{\mcitedefaultmidpunct}
{}{\mcitedefaultseppunct}\relax
\EndOfBibitem
\bibitem[vir(2020)]{viralzone}
2020; Viralzone News, https://viralzone.expasy.org\relax
\mciteBstWouldAddEndPuncttrue
\mciteSetBstMidEndSepPunct{\mcitedefaultmidpunct}
{\mcitedefaultendpunct}{\mcitedefaultseppunct}\relax
\EndOfBibitem
\bibitem[ncb(2020)]{ncbi}
2020; The National Center for Biotechnology Information,
  https://www.ncbi.nlm.nih.gov\relax
\mciteBstWouldAddEndPuncttrue
\mciteSetBstMidEndSepPunct{\mcitedefaultmidpunct}
{\mcitedefaultendpunct}{\mcitedefaultseppunct}\relax
\EndOfBibitem
\bibitem[Shanker \latin{et~al.}(2020)Shanker, Bhanu, and Alluri]{Shanker2020}
Shanker,~A.; Bhanu,~D.; Alluri,~A. {Analysis of Whole Genome Sequences and
  Homology Modelling of a 3C Like Peptidase and a Non-Structural Protein of the
  Novel Coronavirus COVID-19 Shows Protein Ligand Interaction with an
  Aza-Peptide and a Noncovalent Lead Inhibitor with Possible Antiviral
  Properties}. \emph{ChemRxiv} \textbf{2020}, \relax
\mciteBstWouldAddEndPunctfalse
\mciteSetBstMidEndSepPunct{\mcitedefaultmidpunct}
{}{\mcitedefaultseppunct}\relax
\EndOfBibitem
\bibitem[Thiel \latin{et~al.}(2003)Thiel, Ivanov, Putics, Hertzig, Schelle,
  Bayer, Wei{\ss}brich, Snijder, Rabenau, Doerr, Gorbalenya, and
  Ziebuhr]{Thiel2003}
Thiel,~V.; Ivanov,~K.~A.; Putics,~A.; Hertzig,~T.; Schelle,~B.; Bayer,~S.;
  Wei{\ss}brich,~B.; Snijder,~E.~J.; Rabenau,~H.; Doerr,~H.~W.;
  Gorbalenya,~A.~E.; Ziebuhr,~J. Mechanisms and Enzymes Involved in SARS
  Coronavirus Genome Expression. \emph{J. Gen. Virol.} \textbf{2003},
  \emph{84}, 2305--2315\relax
\mciteBstWouldAddEndPuncttrue
\mciteSetBstMidEndSepPunct{\mcitedefaultmidpunct}
{\mcitedefaultendpunct}{\mcitedefaultseppunct}\relax
\EndOfBibitem
\bibitem[Hilgenfeld(2014)]{Hilgenfeld2014}
Hilgenfeld,~R. From SARS to MERS: Crystallographic Studies on Coronaviral
  Proteases Enable Antiviral Drug Design. \emph{FEBS J.} \textbf{2014},
  \emph{281}, 4085--4096\relax
\mciteBstWouldAddEndPuncttrue
\mciteSetBstMidEndSepPunct{\mcitedefaultmidpunct}
{\mcitedefaultendpunct}{\mcitedefaultseppunct}\relax
\EndOfBibitem
\bibitem[Anand \latin{et~al.}(2003)Anand, Ziebuhr, Wadhwani, Mesters, and
  Hilgenfeld]{Anand2003}
Anand,~K.; Ziebuhr,~J.; Wadhwani,~P.; Mesters,~J.~R.; Hilgenfeld,~R.
  Coronavirus Main Proteinase (3CLpro) Structure: Basis for Design of Anti-SARS
  Drugs. \emph{Science} \textbf{2003}, \emph{300}, 1763--1767\relax
\mciteBstWouldAddEndPuncttrue
\mciteSetBstMidEndSepPunct{\mcitedefaultmidpunct}
{\mcitedefaultendpunct}{\mcitedefaultseppunct}\relax
\EndOfBibitem
\bibitem[Chuck \latin{et~al.}(2010)Chuck, Chong, Chen, Chow, Wan, and
  Wong]{Chuck2010}
Chuck,~C.-P.; Chong,~L.-T.; Chen,~C.; Chow,~H.-F.; Wan,~D. C.-C.; Wong,~K.-B.
  Profiling of Substrate Specificity of SARS-CoV 3CL. \emph{PloS One}
  \textbf{2010}, \emph{5}, e13197--e13197\relax
\mciteBstWouldAddEndPuncttrue
\mciteSetBstMidEndSepPunct{\mcitedefaultmidpunct}
{\mcitedefaultendpunct}{\mcitedefaultseppunct}\relax
\EndOfBibitem
\bibitem[Shi \latin{et~al.}(2008)Shi, Sivaraman, and Song]{Shi4620}
Shi,~J.; Sivaraman,~J.; Song,~J. Mechanism for Controlling the Dimer-Monomer
  Switch and Coupling Dimerization to Catalysis of the Severe Acute Respiratory
  Syndrome Coronavirus 3C-Like Protease. \emph{J. Virol.} \textbf{2008},
  \emph{82}, 4620--4629\relax
\mciteBstWouldAddEndPuncttrue
\mciteSetBstMidEndSepPunct{\mcitedefaultmidpunct}
{\mcitedefaultendpunct}{\mcitedefaultseppunct}\relax
\EndOfBibitem
\bibitem[Johansson(2012)]{Johansson2012}
Johansson,~M.~H. Reversible Michael Additions: Covalent Inhibitors and
  Prodrugs. \emph{Mini-Rev. Med. Chem.} \textbf{2012}, \emph{12},
  1330--1344\relax
\mciteBstWouldAddEndPuncttrue
\mciteSetBstMidEndSepPunct{\mcitedefaultmidpunct}
{\mcitedefaultendpunct}{\mcitedefaultseppunct}\relax
\EndOfBibitem
\bibitem[Vasudevan \latin{et~al.}(2019)Vasudevan, Argiriadi, Baranczak,
  Friedman, Gavrilyuk, Hobson, Hulce, Osman, and Wilson]{Vasudevan2019}
Vasudevan,~A.; Argiriadi,~M.~A.; Baranczak,~A.; Friedman,~M.~M.; Gavrilyuk,~J.;
  Hobson,~A.~D.; Hulce,~J.~J.; Osman,~S.; Wilson,~N.~S. In \emph{Chapter One -
  Covalent Binders in Drug Discovery}; Witty,~D.~R., Cox,~B., Eds.; Prog. Med.
  Chem.; Elsevier, 2019; Vol.~58; pp 1 -- 62\relax
\mciteBstWouldAddEndPuncttrue
\mciteSetBstMidEndSepPunct{\mcitedefaultmidpunct}
{\mcitedefaultendpunct}{\mcitedefaultseppunct}\relax
\EndOfBibitem
\bibitem[Yang \latin{et~al.}(2003)Yang, Yang, Ding, Liu, Lou, Zhou, Sun, Mo,
  Ye, Pang, Gao, Anand, Bartlam, Hilgenfeld, and Rao]{Yang2003}
Yang,~H.; Yang,~M.; Ding,~Y.; Liu,~Y.; Lou,~Z.; Zhou,~Z.; Sun,~L.; Mo,~L.;
  Ye,~S.; Pang,~H.; Gao,~G.~F.; Anand,~K.; Bartlam,~M.; Hilgenfeld,~R.; Rao,~Z.
  The Crystal Structures of Severe Acute Respiratory Syndrome Virus Main
  Protease and Its Complex with an Inhibitor. \emph{Proc. Natl. Acad. Sci. USA}
  \textbf{2003}, \emph{100}, 13190--13195\relax
\mciteBstWouldAddEndPuncttrue
\mciteSetBstMidEndSepPunct{\mcitedefaultmidpunct}
{\mcitedefaultendpunct}{\mcitedefaultseppunct}\relax
\EndOfBibitem
\bibitem[Jacobs \latin{et~al.}(2010)Jacobs, Zhou, Dawson, Daniels, Hodder,
  Tokars, Mesecar, Lindsley, , and Stauffer.]{Jacobs2010}
Jacobs,~J.; Zhou,~S.; Dawson,~E.; Daniels,~J.~S.; Hodder,~P.; Tokars,~V.;
  Mesecar,~A.; Lindsley,~C.~W.; ; Stauffer.,~S.~R. Discovery of Non-Covalent
  Inhibitors of the SARS Main Proteinase 3CLpro. \emph{Probe Reports from the
  NIH Molecular Libraries Program} \textbf{2010},
  https://www.ncbi.nlm.nih.gov/books/NBK133447/\relax
\mciteBstWouldAddEndPuncttrue
\mciteSetBstMidEndSepPunct{\mcitedefaultmidpunct}
{\mcitedefaultendpunct}{\mcitedefaultseppunct}\relax
\EndOfBibitem
\bibitem[Jacobs \latin{et~al.}(2013)Jacobs, Grum-Tokars, Zhou, Turlington,
  Saldanha, Chase, Eggler, Dawson, Baez-Santos, Tomar, Mielech, Baker,
  Lindsley, Hodder, Mesecar, and Stauffer]{Jacobs2013}
Jacobs,~J.; Grum-Tokars,~V.; Zhou,~Y.; Turlington,~M.; Saldanha,~S.~A.;
  Chase,~P.; Eggler,~A.; Dawson,~E.~S.; Baez-Santos,~Y.~M.; Tomar,~S.;
  Mielech,~A.~M.; Baker,~S.~C.; Lindsley,~C.~W.; Hodder,~P.; Mesecar,~A.;
  Stauffer,~S.~R. Discovery, Synthesis, and Structure-Based Optimization of a
  Series of N-(tert-Butyl)-2-(N-arylamido)-2-(pyridin-3-yl) Acetamides (ML188)
  as Potent Noncovalent Small Molecule Inhibitors of the Severe Acute
  Respiratory Syndrome Coronavirus (SARS-CoV) 3CL Protease. \emph{J. Med.
  Chem.} \textbf{2013}, \emph{56}, 534--546\relax
\mciteBstWouldAddEndPuncttrue
\mciteSetBstMidEndSepPunct{\mcitedefaultmidpunct}
{\mcitedefaultendpunct}{\mcitedefaultseppunct}\relax
\EndOfBibitem
\bibitem[Liu \latin{et~al.}()Liu, Zhang, Jin, Yang, and Rao]{6lu7}
Liu,~X.; Zhang,~B.; Jin,~Z.; Yang,~H.; Rao,~Z. The Crystal Structure of
  2019-nCoV Main Protease in Complex with an Inhibitor N3. RSCB PDB, pdbode:
  6LU7\relax
\mciteBstWouldAddEndPuncttrue
\mciteSetBstMidEndSepPunct{\mcitedefaultmidpunct}
{\mcitedefaultendpunct}{\mcitedefaultseppunct}\relax
\EndOfBibitem
\bibitem[Hu \latin{et~al.}(2009)Hu, Zhang, Li, Wang, Chen, Chen, Ding, Jiang,
  and Shen]{Hu2009}
Hu,~T.; Zhang,~Y.; Li,~L.; Wang,~K.; Chen,~S.; Chen,~J.; Ding,~J.; Jiang,~H.;
  Shen,~X. Two Adjacent Mutations on the Dimer Interface of SARS Coronavirus
  3C-like Protease Cause Different Conformational Changes in Crystal Structure.
  \emph{Virology} \textbf{2009}, \emph{388}, 324 -- 334\relax
\mciteBstWouldAddEndPuncttrue
\mciteSetBstMidEndSepPunct{\mcitedefaultmidpunct}
{\mcitedefaultendpunct}{\mcitedefaultseppunct}\relax
\EndOfBibitem
\bibitem[pla()]{play}
PlayMolecule\texttrademark, https://www.acellera.com, accessed 20 February
  2020\relax
\mciteBstWouldAddEndPuncttrue
\mciteSetBstMidEndSepPunct{\mcitedefaultmidpunct}
{\mcitedefaultendpunct}{\mcitedefaultseppunct}\relax
\EndOfBibitem
\bibitem[Skalic \latin{et~al.}(2019)Skalic, Sabbadin, Sattarov, Sciabola, and
  De~Fabritiis]{skalic2019target}
Skalic,~M.; Sabbadin,~D.; Sattarov,~B.; Sciabola,~S.; De~Fabritiis,~G. From
  Target to Drug: Generative Modeling for the Multimodal Structure-Based Ligand
  Design. \emph{Mol. Pharm.} \textbf{2019}, \emph{16}, 4282--4291\relax
\mciteBstWouldAddEndPuncttrue
\mciteSetBstMidEndSepPunct{\mcitedefaultmidpunct}
{\mcitedefaultendpunct}{\mcitedefaultseppunct}\relax
\EndOfBibitem
\bibitem[Morris \latin{et~al.}(2009)Morris, Huey, Lindstrom, Sanner, Belew,
  Goodsell, and Olson]{Autodock}
Morris,~G.~M.; Huey,~R.; Lindstrom,~W.; Sanner,~M.~F.; Belew,~R.~K.;
  Goodsell,~D.~S.; Olson,~A.~J. AutoDock4 and AutoDockTools4: Automated Docking
  with Selective Receptor Flexibility. \emph{J. Comput. Chem.} \textbf{2009},
  \emph{30}, 2785--2791\relax
\mciteBstWouldAddEndPuncttrue
\mciteSetBstMidEndSepPunct{\mcitedefaultmidpunct}
{\mcitedefaultendpunct}{\mcitedefaultseppunct}\relax
\EndOfBibitem
\bibitem[O'Boyle \latin{et~al.}(2011)O'Boyle, Banck, James, Morley,
  Vandermeersch, and Hutchison]{Babel}
O'Boyle,~N.~M.; Banck,~M.; James,~C.~A.; Morley,~C.; Vandermeersch,~T.;
  Hutchison,~G.~R. Open Babel: An Open Chemical Toolbox. \emph{J. Cheminf.}
  \textbf{2011}, \emph{3}, 33\relax
\mciteBstWouldAddEndPuncttrue
\mciteSetBstMidEndSepPunct{\mcitedefaultmidpunct}
{\mcitedefaultendpunct}{\mcitedefaultseppunct}\relax
\EndOfBibitem
\bibitem[Cheng \latin{et~al.}(2007)Cheng, Zhao, Li, Lin, Xu, Zhang, Li, Wang,
  and Lai]{Cheng2007}
Cheng,~T.; Zhao,~Y.; Li,~X.; Lin,~F.; Xu,~Y.; Zhang,~X.; Li,~Y.; Wang,~R.;
  Lai,~L. Computation of Octanol-Water Partition Coefficients by Guiding an
  Additive Model with Knowledge. \emph{J. Chem. Inf. Model.} \textbf{2007},
  \emph{47}, 2140--2148\relax
\mciteBstWouldAddEndPuncttrue
\mciteSetBstMidEndSepPunct{\mcitedefaultmidpunct}
{\mcitedefaultendpunct}{\mcitedefaultseppunct}\relax
\EndOfBibitem
\bibitem[Graziano \latin{et~al.}(2006)Graziano, McGrath, Yang, and
  Mangel]{Graziano2006}
Graziano,~V.; McGrath,~W.~J.; Yang,~L.; Mangel,~W.~F. SARS CoV Main Proteinase:
  The Monomer-Dimer Equilibrium Dissociation Constant. \emph{Biochemistry}
  \textbf{2006}, \emph{45}, 14632--14641\relax
\mciteBstWouldAddEndPuncttrue
\mciteSetBstMidEndSepPunct{\mcitedefaultmidpunct}
{\mcitedefaultendpunct}{\mcitedefaultseppunct}\relax
\EndOfBibitem
\bibitem[Kim \latin{et~al.}(2016)Kim, Thiessen, Bolton, Chen, Fu, Gindulyte,
  Han, He, He, Shoemaker, Wang, Yu, Zhang, and Bryant]{pubchem}
Kim,~S.; Thiessen,~P.~A.; Bolton,~E.~E.; Chen,~J.; Fu,~G.; Gindulyte,~A.;
  Han,~L.; He,~J.; He,~S.; Shoemaker,~B.~A.; Wang,~J.; Yu,~B.; Zhang,~J.;
  Bryant,~S.~H. PubChem Substance and Compound Databases. \emph{Nucleic Acids
  Res.} \textbf{2016}, \emph{44}, D1202--D1213\relax
\mciteBstWouldAddEndPuncttrue
\mciteSetBstMidEndSepPunct{\mcitedefaultmidpunct}
{\mcitedefaultendpunct}{\mcitedefaultseppunct}\relax
\EndOfBibitem
\bibitem[Pronk \latin{et~al.}(2013)Pronk, P\'all, Schulz, Larsson, Bjelkmar,
  Apostolov, Shirts, Smith, Kasson, van~der Spoel, Hess, and Lindahl]{gromacs}
Pronk,~S.; P\'all,~S.; Schulz,~R.; Larsson,~P.; Bjelkmar,~P.; Apostolov,~R.;
  Shirts,~M.~R.; Smith,~J.~C.; Kasson,~P.~M.; van~der Spoel,~D.; Hess,~B.;
  Lindahl,~E. GROMACS 4.5: a High-Throughput and Highly Parallel Open Source
  Molecular Simulation Toolkit. \emph{Bioinformatics} \textbf{2013}, \emph{29},
  845\relax
\mciteBstWouldAddEndPuncttrue
\mciteSetBstMidEndSepPunct{\mcitedefaultmidpunct}
{\mcitedefaultendpunct}{\mcitedefaultseppunct}\relax
\EndOfBibitem
\bibitem[Van Der~Spoel \latin{et~al.}(2005)Van Der~Spoel, Lindahl, Hess,
  Groenhof, Mark, and Berendsen]{gromacs1}
Van Der~Spoel,~D.; Lindahl,~E.; Hess,~B.; Groenhof,~G.; Mark,~A.~E.;
  Berendsen,~H. J.~C. GROMACS: Fast, Flexible, and Free. \emph{J. Comput.
  Chem.} \textbf{2005}, \emph{26}, 1701--1718\relax
\mciteBstWouldAddEndPuncttrue
\mciteSetBstMidEndSepPunct{\mcitedefaultmidpunct}
{\mcitedefaultendpunct}{\mcitedefaultseppunct}\relax
\EndOfBibitem
\bibitem[Macchiagodena \latin{et~al.}(2019)Macchiagodena, Pagliai, Andreini,
  Rosato, and Procacci]{macchiagodena2019}
Macchiagodena,~M.; Pagliai,~M.; Andreini,~C.; Rosato,~A.; Procacci,~P.
  Upgrading and Validation of the AMBER Force Field for Histidine and Cysteine
  Zinc(II)-Binding Residues in Sites with Four Protein Ligands. \emph{J. Chem.
  Inf. Model.} \textbf{2019}, \emph{59}, 3803--3816\relax
\mciteBstWouldAddEndPuncttrue
\mciteSetBstMidEndSepPunct{\mcitedefaultmidpunct}
{\mcitedefaultendpunct}{\mcitedefaultseppunct}\relax
\EndOfBibitem
\bibitem[Procacci and Chelli(2017)Procacci, and Chelli]{statbind}
Procacci,~P.; Chelli,~R. {Statistical Mechanics of Ligand-Receptor Noncovalent
  Association, Revisited: Binding Site and Standard State Volumes in Modern
  Alchemical Theories}. \emph{J. Chem. Theory Comput.} \textbf{2017},
  \emph{13}, 1924--1933\relax
\mciteBstWouldAddEndPuncttrue
\mciteSetBstMidEndSepPunct{\mcitedefaultmidpunct}
{\mcitedefaultendpunct}{\mcitedefaultseppunct}\relax
\EndOfBibitem
\bibitem[Gilson \latin{et~al.}(1997)Gilson, Given, Bush, and
  McCammon]{Gilson1997}
Gilson,~M.~K.; Given,~J.~A.; Bush,~B.~L.; McCammon,~J.~A. The
  Statistical-Thermodynamic Basis for Computation of Binding Affinities: A
  Critical Review. \emph{Biophys. J.} \textbf{1997}, \emph{72},
  1047--1069\relax
\mciteBstWouldAddEndPuncttrue
\mciteSetBstMidEndSepPunct{\mcitedefaultmidpunct}
{\mcitedefaultendpunct}{\mcitedefaultseppunct}\relax
\EndOfBibitem
\bibitem[Nerattini \latin{et~al.}(2016)Nerattini, Chelli, and
  Procacci]{procaccipccp2016a}
Nerattini,~F.; Chelli,~R.; Procacci,~P. II. Dissociation Free Energies in
  Drug–Receptor Systems Via Nonequilibrium Alchemical Simulations:
  Application to the FK506-Related Immunophilin Ligands. \emph{Phys. Chem.
  Chem. Phys.} \textbf{2016}, \emph{18}, 15005--15018\relax
\mciteBstWouldAddEndPuncttrue
\mciteSetBstMidEndSepPunct{\mcitedefaultmidpunct}
{\mcitedefaultendpunct}{\mcitedefaultseppunct}\relax
\EndOfBibitem
\bibitem[Procacci(2016)]{procaccipccp2016}
Procacci,~P. I. Dissociation Free Energies of Drug-Receptor Systems Via
  Non-equilibrium Alchemical Simulations: a Theoretical Framework. \emph{Phys.
  Chem. Chem. Phys.} \textbf{2016}, \emph{18}, 14991--15004\relax
\mciteBstWouldAddEndPuncttrue
\mciteSetBstMidEndSepPunct{\mcitedefaultmidpunct}
{\mcitedefaultendpunct}{\mcitedefaultseppunct}\relax
\EndOfBibitem
\bibitem[Shirts and Mobley(2013)Shirts, and Mobley]{Shirts2013}
Shirts,~M.~R.; Mobley,~D.~L. An Introduction to Best Practices in Free Energy
  Calculations. \emph{Methods Mol. Biol.} \textbf{2013}, \emph{924},
  271--311\relax
\mciteBstWouldAddEndPuncttrue
\mciteSetBstMidEndSepPunct{\mcitedefaultmidpunct}
{\mcitedefaultendpunct}{\mcitedefaultseppunct}\relax
\EndOfBibitem
\end{mcitethebibliography}
\end{document}